**Electron dynamics and $SiO_2$ etching profile evolution in capacitive Ar/$CHF_3$ discharges driven by sawtooth-tailored voltage waveforms**

Wan Dong[1,2], Liu-Qin Song[1], Yi-Fan Zhang[1], Li Wang[2], Yuan-Hong Song[1*], and Julian Schulze[2]

[1]Key Laboratory of Materials Modification by Laser, Ion and Electron Beams (Ministry of Education), School of Physics, Dalian University of Technology, Dalian 116024, People's Republic of China

[2]Chair of Applied Electrodynamics and Plasma Technology, Department of Electrical Engineering and Information Science, Ruhr-University Bochum, D-44780, Bochum, Germany

[*]E-mail: songyh@dlut.edu.cn

**Abstract**

The electron dynamics and $SiO_2$ etching profile evolution in capacitively coupled Ar/$CHF_3$ plasmas driven by sawtooth-waveforms are investigated based on a one-dimensional fluid/Monte-Carlo (MC) model coupled with an etching profile evolution model. The effects of the sawtooth-waveforms synthesized from different numbers of consecutive harmonics, *N*, of a fundamental frequency of 13.56 MHz on the electron dynamics, ion and neutral transport, as well as the etching profile evolution are revealed in different mixtures of Ar/$CHF_3$. By increasing *N*, a reduction in electronegativity, a decrease of the DC self-bias voltage, and a transition of the discharge mode from the Drift-Ambipolar (DA) to an α-DA hybrid mode is observed accompanied by an enhanced plasma asymmetry. As the $CHF_3$ gas admixture increases, the electronegativity initially increases and then decreases, following a similar trend as the absolute value of the DC self-bias voltage. This is mainly caused by the change in ionization, attachment and de-attachment reaction rates. The obtained results show that placing the substrate on the grounded electrode and using a higher number of harmonic frequencies (*N*) can achieve a faster etching rate, since higher ion fluxes can be obtained in these scenarios. Additionally, the Ar/$CHF_3$ gas mixing ratio impacts the neutral surface coverage, which in turn affects the etching rate. Therefore, selecting an appropriate gas mixture is also essential for optimizing etching results.

## 1 Introduction

Radio frequency capacitively coupled plasmas (RF CCPs) are often used for applications such as plasma enhanced chemical vapor deposition (PECVD) and plasma etching in semiconductor manufacturing [1,2]. Complex (reactive and electronegative) gas mixtures [3,4], such as Ar/$C_4F_8$ are typically used in such applications [5,6] to optimize the ratio of the ion and radical fluxes and methods to precisely control their energy distribution functions at the wafer are required to reach a good process quality. With the continuous decrease of the feature size and increase of the wafer area, development of the next-generation etching devices becomes necessary while the traditional CCPs driven by single/dual frequency driving voltage waveforms no longer provide sufficient process control. Several researchers have revealed the big advantages of tailored voltage waveforms (TVWs) synthesized from a fundamental frequency and its higher harmonics with adjustable harmonics' amplitudes and phase shifts for the control of electron heating dynamics [16-18], neutral radical generation rates [19], ion flux and energy distribution functions [7-13], as well as the suppression of plasma non-uniformities caused by electromagnetic effects [14,15]. For instance, in CCPs driven by TVWs, the DC self-bias and asymmetries of the discharge can be controlled by adjusting the driving voltage waveform, which is known as the Electrical Asymmetry Effect (EAE) [20-22]. The value of the DC self-bias can be adjusted by tuning the phase shift between the different harmonics, which provides an important way to control the ion energy independently from the ion flux [23-25]. Depending on the shape of the TVWs, the EAE is typically classified into two types, i.e. the Amplitude Asymmetry Effect (AAE) [26-28], caused by the different absolute values of the positive and negative extrema of the driving voltage waveform, and the Slope Asymmetry Effect (SAE) caused by the different rising and falling slopes of the driving voltage waveform [29-31].

A series of studies on the SAE and AAE in terms of the control of electron heating, discharge mode, ion energy and flux were conducted previously [32-34]. Drastically different SAEs can be generated in electropositive and electronegative discharges driven by sawtooth voltage waveforms due to their different operation mechanisms, i.e. electropositive discharges are typically operated in the α mode, where electrons are mostly accelerated by the expanding RF sheaths, while strongly electronegative discharges are usually operated in the Drift-Ambipolar (DA) mode where electrons are accelerated by strong drift-ambipolar electric fields in the plasma bulk and near the collapsing sheath edges. The presence of these different modes of plasma operation in combination with the SAE can result in an opposite sign of the DC self-bias, which can be positive and negative in electropositive and electronegative discharges, respectively, that are otherwise symmetric [35]. Electric field reversals were found to be generated during the sheath collapse in the presence of the AAE. Such field reversals can accelerate electrons toward the wafer, enabling them to penetrate deeply into high aspect ratio etch (HARE) features. This can play a crucial role in preventing positive surface charging within the etch features, which would otherwise lead to profile distortions and reduced etch rates [36, 37].

Despite the importance of these fundamental research results obtained in the past years on the topic of controlling plasma properties through TVWs, their ultimate effects on applications, e.g. plasma etching processes, are still not fully understood. In addition, most of the previous studies have been focused on single gas discharges, such as Ar, $O_2$, $H_2$, $Cl_2$, $CF_4$, $SiH_4$, with only a few investigations of their mixtures [12, 36-40]. Very limited investigations have been performed in $CHF_3$ discharges driven by TVWs, although $CHF_3$ discharges have been demonstrated to have significant advantages in etching [41-45]. For instance, Metzler et al. found that $CHF_3$ can provide better deposition of fluorocarbon (FC) polymer compared to $C_4F_8$ during atomic layer etching (ALE) of $SiO_2$ and Si, which enables selective $SiO_2$/Si etching [43]. In addition, the effects of the high driving frequency (HF) on the etching of SiCOH films with a low dielectric constant (low-k) were investigated in dual-frequency driven $CHF_3$ CCPs [46]. It was found that increasing the HF reduces the power threshold required to suppress the FC deposition on the film surface, which enhances the etching of SiCOH films, since both the ion energy and F atom concentration are increased. Moreover, a significant dependence of the electron and neutral radical densities, e.g. the $CF_2$ density, on the various discharge parameters, i.e. high/low frequency power, gas pressure, electrode gap, gas mixing ratio etc., were revealed by Liu et al [47] in dual-frequency $CHF_3$/Ar CCPs. Further studies were also conducted on reactive ion etching of GaN in $CHF_3$/Ar gas mixtures, which revealed a linear increase of the etch rate as a function of the RF power and a decrease of it as a function of pressure [48]. ALE of TiN was demonstrated to be achieved based on $CHF_3/O_2$ downstream plasma exposure followed by infrared irradiation [49], which is promising for the atomic-scale fabrication of next-generation 3D devices.

Overall, previous studies have shown that $CHF_3$ discharges play important roles in both the etching of conventional semiconductors such as Si and the etching of third-generation semiconductors such as GaN, SiC, and $Ga_2O_3$. However, previous investigations were mostly restricted to single/dual frequency discharges, which have limited flexibilities regarding the control of particle distributions as well as the etching process, without TVWs being applied. Thus, in this work, we study the discharge characteristics and the etching evolution in Ar/$CHF_3$ CCPs driven by TVWs synthesized from different numbers of consecutive harmonics, $N$, of a fundamental driving frequency of 13.56 MHz based on a one-dimensional fluid/Monte-Carlo (MC) hybrid model coupled with an etching profile evolution model. The etching of patterned $SiO_2$ wafers is considered. The effects of the gas mixing ratio on the discharge and etching process are also investigated. The paper is structured in the following way: a description of the computational model is presented in Section 2. The simulation results are discussed in Section 3, and finally, the conclusions are given in section 4.

## 2 Model description

Simulations are conducted in Ar/$CHF_3$ CCPs based on a one-dimensional (1D) fluid/MC hybrid model coupled with an etching profile evolution model. Introductions to both models are provided below.

### 2.1 One-dimensional fluid/MC hybrid model

The hybrid plasma model comsists of a one-dimensional fluid model along with electron and ion MC models. The fluid model simulates the transport of electrons, ions, and neutral radicals in the following way [50, 51]: continuity equations are solved for electrons, ions, and neutrals, the ion transport is described by the ion momentum balance equation, while the drift-diffusion approximation is employed for electrons, the transport of neutral radicals is described by the diffusion equations. The following particle species are included in the model: $Ar^+$, $CF_3^+$, $CHF_2^+$, $CF_2^+$, $CHF^+$, $CF^+$, $CH^+$, $F^+$, $F^-$, $Ar^*$, H, F, HF, $F_2$, $CHF_2$, $CF_3$, $CF_2$, CHF, CF. The diffusion coefficients $D_n$ of the neutrals are calculated based on the expression given in Ref. [50] with the binary collision diameters ($\sigma_i$) and potential energies ($\varepsilon_i/k_b$) taken from the references listed in table 1. The interaction of neutral particles with boundary surfaces is determined by the surface sticking coefficients listed in table 1 separately for each neutral species. The spatio-temporal distributions of the potential and the electrostatic field are obtained by solving Poisson's equation. Based on the cold fluid approximation, the ion and neutral temperature is assumed to be constant at 300 K in this work, while the electron energy distribution and temperature are obtained from the electron MC model.

**Table 1**. Lennard-Jones parameters including binary collision diameters ($\sigma_i$) as well as potential energies ($\varepsilon_i/k_b$), and sticking coefficient ($s_j$) of the neutral particles.

| Neutrals | $\sigma_i$(Å) | $\varepsilon_i/k_b$(K) | $s_j$ |
|---|---|---|---|
| $Ar^*$ | 3.54[52] | 93.3[52] | 0.8[53] |
| $CF_3$ | 4.32[52] | 121[52] | 0.017[53] |
| $CHF_2$ | 3.98[52] | 214[52] | 0.005[55] |
| $CF_2$ | 3.977[52] | 108[52] | 0.02[53] |
| CHF | 3.645[52] | 210.2[52] | 0.02[55] |
| CF | 3.635[52] | 94.2[52] | 0.036[53] |
| F | 2.968[52] | 112.6[52] | 0.02[53] |
| H | 2.708[52] | 37[52] | 0.9[54] |
| HF | 3.148[52] | 330[52] | 0.001[55] |
| $F_2$ | 3.61[52] | 121[52] | 0.02[51] |

In the electron MC model, 25 electron-neutral impact reactions, including elastic collisions, ionization, excitation, and dissociative attachment, are considered, as listed in table A1 in the Appendix [56-58]. The cross sections used in this model are taken from the references listed in the last column of this table. The spatio-temporal electron energy distribution function (EEDF) is obtained statistically from the electron MC model. The kinetic effects are included in this way, which cannot be captured by pure fluid models. Based on the EEDF, the electron temperature and electron-neutral collision rate coefficients are calculated. In addition to the electron-neutral reactions, 58 reactions between electrons and ions, ions and ions, ions and neutrals, as well as neutrals and neutrals are considered, as listed in table A2 in the Appendix. The corresponding rate coefficients are also listed, which are given based on the references listed in the last column in table A2 [47, 59-63]. The ion energy distribution functions (IEDFs) at the electrode are obtained from the ion MC model, where the ions accelerated in the sheaths, the elastic and charge exchange collisions between ions and neutrals (Ar and $CHF_3$) are considered self-consistently. Langevin cross sections are used for the ion-neutral collisions.

The above models are coupled in the following way: the fluid model provides the electric field as well as the electron, ion, and neutral densities along with their fluid velocities, which are passed to the electron MC model after each RF (13.56 MHz) cycle.

The collision reaction rate coefficients and mean electron energy are then calculated in the electron MC model based on the EEDF and are transferred back to the fluid model for the next cycle. Convergence is typically achieved after approximately 40,000 RF cycles. The ion MC model is executed during the last 100 RF cycles based on the spatio-temporal electric field obtained from the fluid model to calculate the IEDFs at the electrode. More details regarding this hybrid model can be found in our previous work [45]. The code has been benchmarked against other published results [55], as shown in Appendix Benchmark (figure A).

Sawtooth-waveforms $V(t)|_{x=0} = -V_0 \sum_{k=1}^{N} \frac{1}{k} sin(2\pi f t)$, with $V_0$= 500 V, $f$ = 13.56 MHz, and $N$ ranging from 1 to 3 (see figure 1), are applied to the powered (bottom) electrode (see figure 2). The Ar/$CHF_3$ gas mixing ratio is varied from 90/10, 70/30, 50/50, 30/70, to 10/90 in this work with the gas pressure being kept at a constant value of 100 mTorr, which is a typical pressure used for etching [64, 65]. The electrode gap is 3 cm. Secondary electron emissions from the electrodes is neglected in this work due to their minor effects under the discharge conditions studied (relatively low pressure) [66].

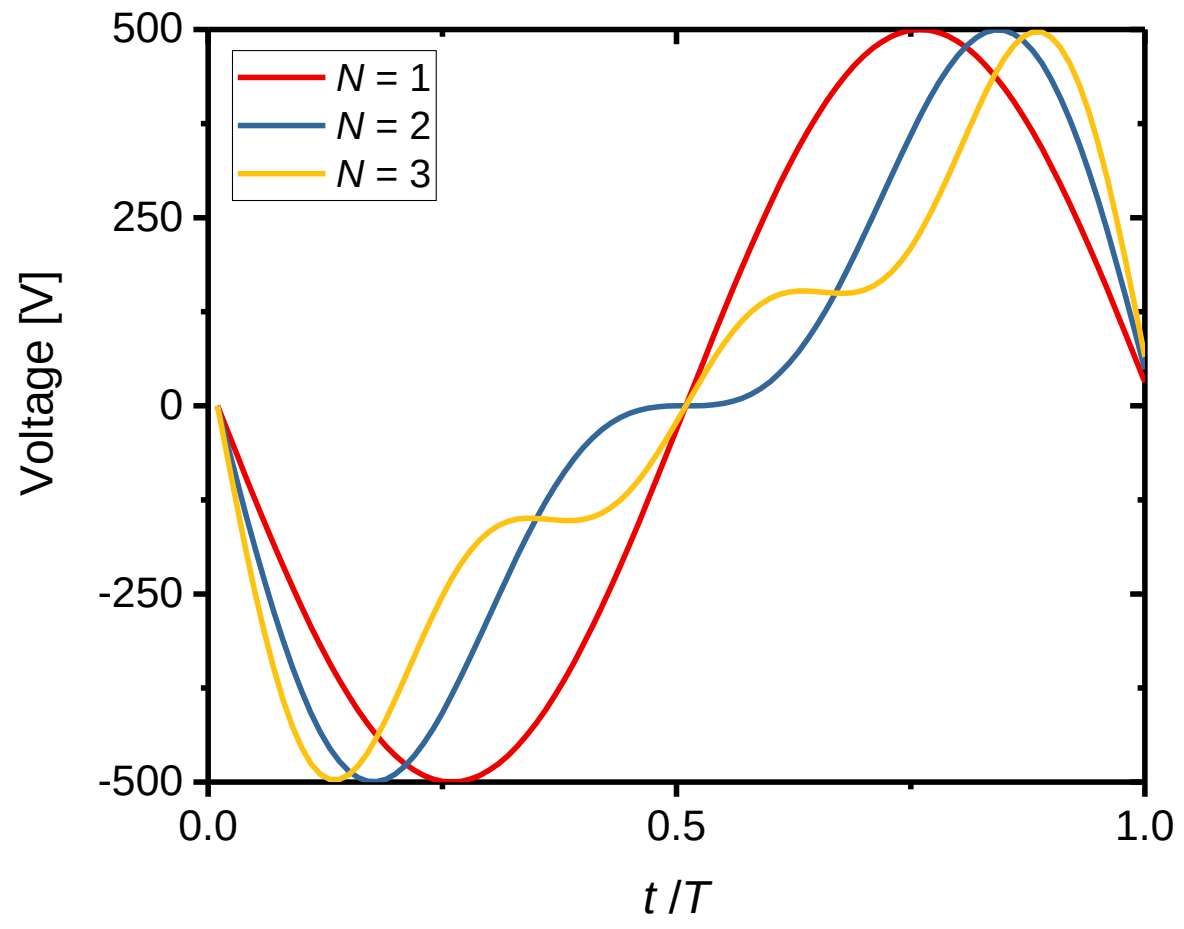

**Figure 1** Single frequency ($N$ = 1) and dual/triple ($N$ = 2 and 3) frequency sawtooth-up waveforms used in this work.

## 2.2 The etching profile evolution model

After convergence of the hybrid model is reached, the ion and neutral fluxes as well as the IEDFs obtained from the hybrid model are used as input parameters in the etching profile evolution model, which includes a surface MC module, a particle tracking module, and a charging effect module. The initial superparticle number of electrons in the etching profile evolution model is set to be thesame as that of ions. Based on the discharge conditions in this work, the velocity distribution of electrons is similar to the Maxwell velocity distribution. Therefore, the velocities of the electrons that enter the etching profile follow a Maxwell velocity distribution and their incident angles are selected randomly.

According to the different materials, the whole etching region with a height of 80 nm and a witdh of 20 nm is divided into distinct zones composed of certain numbers of cells based on the cellular method, as indicated by different colors in the zoom-in plot in figure 2. Each cell has a length of 0.2 nm and represents one atom/molecule of the corresponding material. The cell will be blank if there is no material. The feature profile is composed of 400 cells in the $x$ direction and 100 cells in the $y$ direction. The gray and black regions in this etch profile correspond to the photoresist (20 nm in thickness) and $SiO_2$ material (50 nm in thickness), respectively. The white part corresponds to the region where no material is present. The initial notch width at the photoresist is 10 nm.

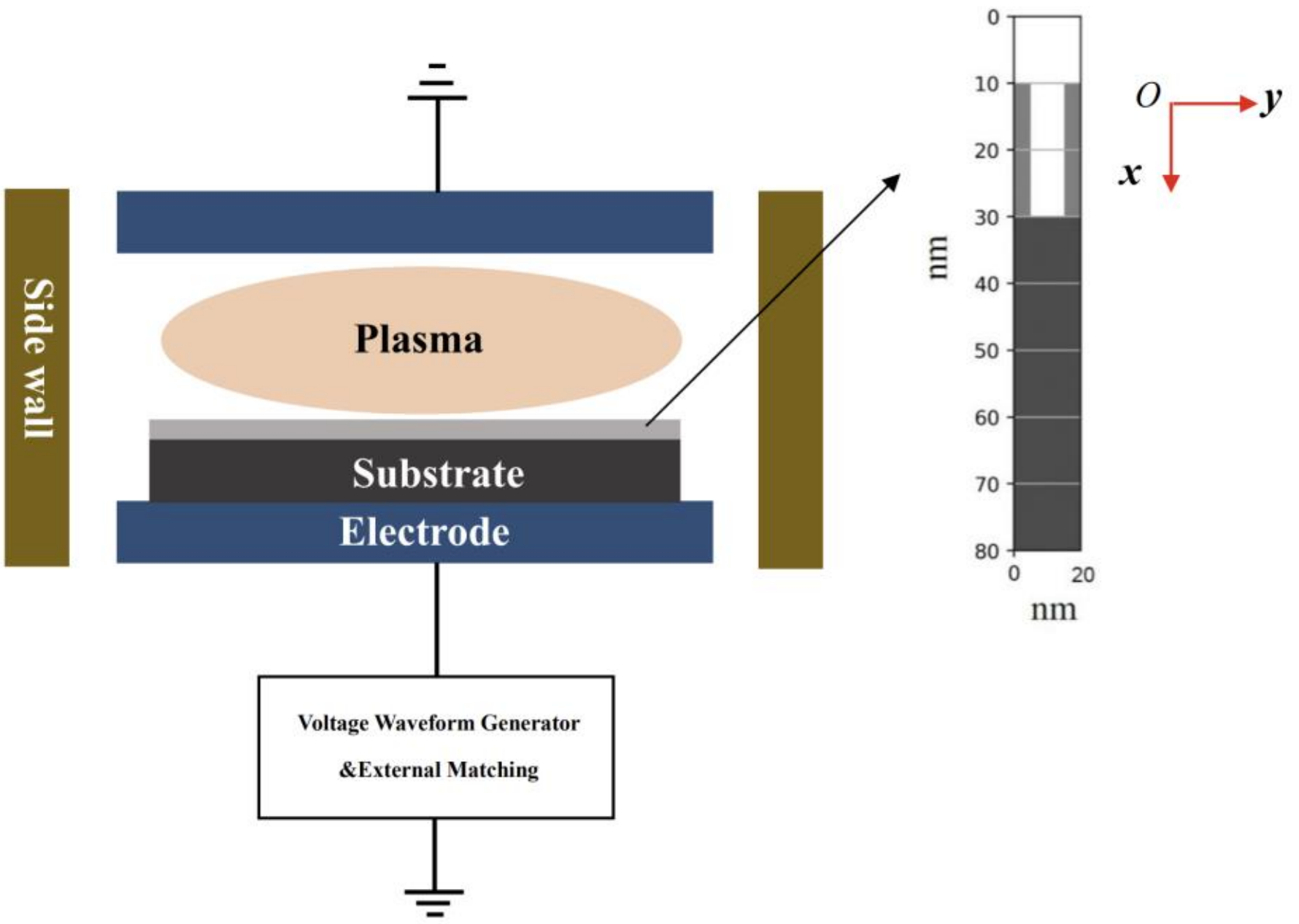

**Figure 2** The schematic diagram of symmetric chamber and initial etching profile used in this work.

The superparticles representing, electrons, ions, and neutral radicals are uniformily distributed in the $y$-direction at $x$ = 0 nm at the beginning of the simulation. These particles are traced in the model until they react with the wafer material. The interactions of these particles with the wafer material is determined by the reflection conditions describd in detail in [59]. For example, the ions will be reflected if their incident angles are larger than 60°. In the case of reflection, the velocity of the particle is updated based on the specular reflection theory. If the reflection condition is not satisfied, a surface reation will occur, the particle is removed from the model and the etching profile is updated by altering its cellular properties. Surface reactions of 12 different types of materials, i.e. Si, $SiO_2$, polymer ($CF_{x(s)}$), ion activated polymer ($CF_x^*{}_{(s)}$), passivation layer of $SiO_2$ ($SiO_2C_xF_{y(s)}$, $SiOC_xF_{y(s)}$, and $X_{SiF(s)}$), passivation layer of Si ($SiF_{(s)}$, $SiF_{2(s)}$, and $SiF_{3(s)}$), hydrogenated polymers ($HP_{(s)}$), and photoresist ($R_{(s)}$) are considered in the model, as shown in table A3 in the Appendix. The corresponding threshold energies, reference energies (which play a calibrating role in sputtering formulas and are used to standardize and adjust the energy relationship between the ions and the surface material molecules), as well as the energy- and angle-dependent reaction probabilities are also listed in the table. Surface charging caused by electrons and ions can change the electric field inside the trench. Such process is considerd in the model based on the charging effect model described in ref. [68]. Based on this model, the etch rates and the time evolution of the etching profiles under various discharge conditions and based on the ion/neutral fluxes and mean ion energies at both electrodes are compared in this work to identify the better etching conditions.

## 3 Results

The effects of the number of harmonics used to synthesize the sawtooth-waveform and the gas mixing ratio on the electron dynamics/etching profile evolution are discussed in the following sections 3.1 and 3.2, respectively.

### 3.1 Effects of the number of harmonics of the sawtooth-waveform on the electron dynamics and etching profiles

Figure 3 presents the spatio-temporal distributions of the electron density, $F^-$ density, electric field, electron power absorption rate, and total ionization rate at different number of harmonics ($N$) of the sawtooth-waveform. The gas pressure is fixed at 100 mTorr and the Ar/$CHF_3$ gas mixing ratio is 70/30. The corresponding driving voltage waveforms as well as the electron current density at the discharge center are given in figure 3 (f1)-(f3). The total ionization rate is the sum of the reaction rates of R1, R4, R19-R25 in table A1, and R31-R35 in table A2. As shown in figures 3 (a1)-(a3), the electron density is increased as a function of $N$, especially in the bulk region. This increase is caused by the enhanced electron power absorption, the spatio-temporally averaged value of which is increased from $1.95\times10^4$ W/m$^2$ ($N$ = 1) to $2.07\times10^4$ W/m$^2$ ($N$ = 2) and $2.30\times10^4$ W/m$^2$ ($N$ = 3). As a result, the $F^-$ density in the bulk region is also increased due to the enhanced dissociative electron attachment (R10 in table A1) and electron impact dissociative ionization (R11 in table A1).

At low values of $N$, the discharge is operated in the Drift-Ambipolar (DA) mode, where electrons are accelerated by the strong drift electric field in the bulk region and by the ambipolar electric field near the collapsing sheath edges [69]. The drift electric field is caused by the low electron density and low plasma conductivity in the bulk region and the ambipolar field is caused by the large electron density gradient generated from the electron density maximum at the sheaths edges (see figure 3 (a1)-(a3)). Since the discharge is symmetric at $N$ = 1 (the DC self-bias is 0 V), same discharge characteristics are found near the top and powered electrodes.

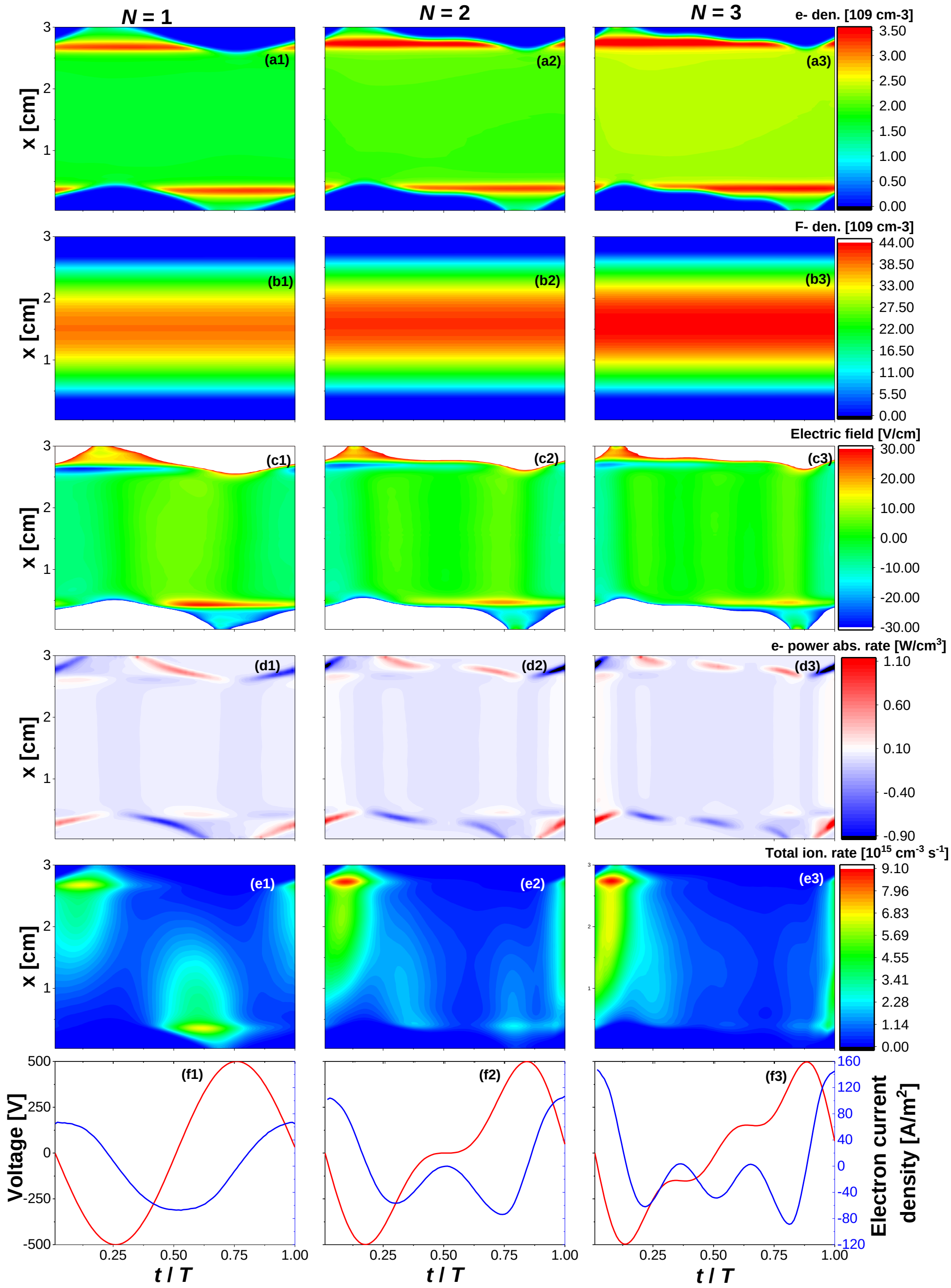

**Figure 3** Spatio-temporal evolutions of the electron density (a1)-(a3), $F^-$ density (b1)-(b3), electric field (c1)-(c3), electron power absorption rate (d1)-(d3), total ionization rate (e1)-(e2) and the driving voltage waveform as well as electron current density at the discharge center (f1)-(f3) at 100 mTorr with the $N$ = 1~3 consecutive harmonics used to synthesize the driving voltage waveform. Discharge conditions: $CHF_3$/Ar (30/70) gas mixture, an electrode gap of 3 cm, and a peak-to-peak driving voltage of 500 V are used.

At higher values of $N$, the sheath at the powered electrode expands much faster compared to the sheath expansion at the grounded electrode. Consequently, a strong electron current is formed during the fast sheath expansion at the powered electrode at around $t$ = 0.0 T (where T is the duration of the fundamental RF period), when the electrons are accelerated upwards, as shown in figure 3 (f2) and (f3). The electron current is lower, when the sheath at the grounded electrode expands, which is maintained for a longer fraction of the fundamental RF period, as shown in figure 3 (f). To drive the high current through the discharge, strong drift electric field is generated, leading to significant electron power absorption within the plasma bulk region. Strong electron power absorption is also formed due to the ambipolar electric field at the collapsing sheath edge near the grounded electrode. In addition, due to the fast sheath expansion at the powered electrode at $N$ = 2 and $N$ = 3 at around $t$ = 0.0 T, the electron heating during the sheath expansion is enhanced, which induces a DA- to α-DA mode transition. When the sheath at the grounded electrode expands, weaker electron power absorption is observed due to the lower current at this time within the fundamental RF period. This induces spatially asymmetric electron power absorption, ionization dynamics, and ion density. A non-zero DC self bias of −64.88 V and −78.97 V is generated at $N$ = 2 and $N$ = 3, respectively, the absolute value of which increases as a function of $N$.

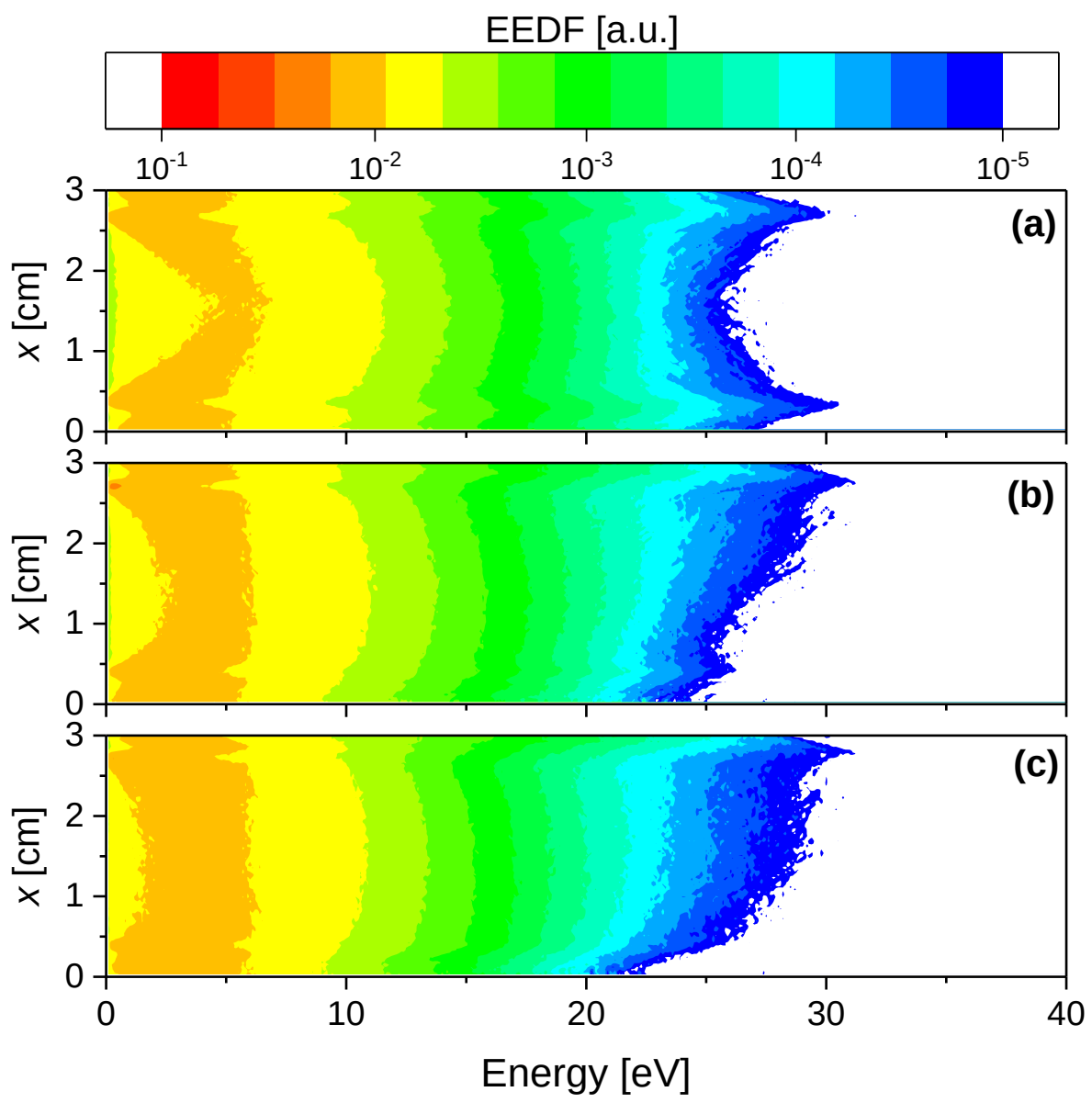

**Figure 4** Spatially resolved and time averaged EEDF for (a) $N$ =1, (b) $N$ = 2, (c) $N$ = 3 at 100 mTorr. The discharge conditions are the same as those in figure 3.

The enhanced electron power absorption leads to a higher electron density at higher values of $N$. As the electron density increases faster as a function of $N$ compared to the negative ion density, the space-time averaged electronegativity decreases from 13.20 ($N$ = 1) to 12.11 ($N$ = 2) and 11.20 ($N$ = 3). Figure 4 shows the space resolved and time averaged EEDF at different values of $N$. As $N$ increases, the fraction of high-energy electrons increases significantly, especially in the vicinity of the grounded electrode. The enhanced high energy tail of the EEDF causes a strong increase of the ionization rate close to the grounded electrode, as shown in figure 3 (e).

The normalized fluxes of the main ion species, i.e. $Ar^+$, $CF_3^+$, $CHF_2^+$, and neutrals, i.e. $CF_2$, CF, F, H at both the powered and grounded electrodes in the cases of different $N$ are presented in figure 5. The neutral fluxes increase as the function of $N$ at both electrodes, while the ion fluxes first decrease and then increase at the powered electrode and continuously increase at the grounded electrode. This trend of the ion fluxes can be explained by the variation of the spatio-temporal ionization dynamics shown in figures 3 (e1)-(e3), which is a result of the enhanced slope asymmetry of the discharge. For $N$ = 1, the spatio-temporal ionization rate is symmetric with two moderately strong maxima appearing near the top and powered electrodes during the respective sheath collapse.

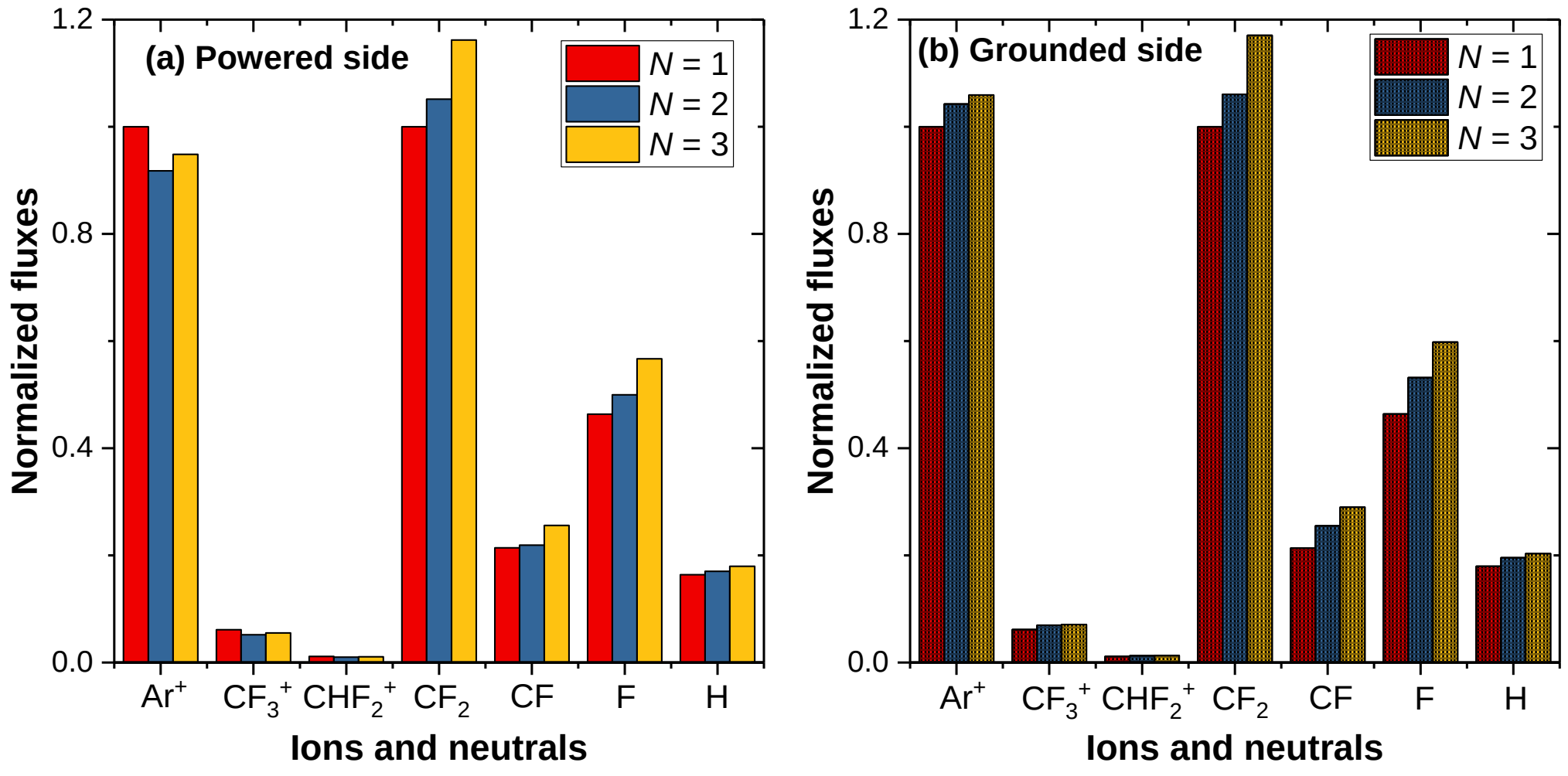

**Figure 5** Normalized fluxes of $Ar^+$, $CF_3^+$, $CHF_2^+$, $CF_2$, CF, F, H, at the (a) powered electrode and the (b) grounded electrode for $N$ = 1, 2, 3. The ion and neutral fluxes are normalized by the $Ar^+$ flux and $CF_2$ flux at $N$ = 1, which are $2.17\times10^{15}$ $cm^{-2}s^{-1}$ and $1.42\times10^{17}$ $cm^{-2}s^{-1}$, respectively. The discharge conditions are the same as those in figure 3.

At $N$ = 2, this symmetry is broken. The ionization close to the grounded electrode is enhanced, while the ionization close to the powered electrode is much lower. As the motion of the ions is mainly determined by their reaction to the local time averaged electric field, ions generated close to one electrode are accelerated towards this electrode by the period-averaged local electric field and would not transport to the opposite electrode. Thus, the ion fluxes to a given electrode are mostly determined by the generation rates of the ions close to this electrode through ionization. Therefore, the less ionization near the powered electrode at $N$ = 2 leads to lower ion fluxes at the powered electrode compared to the case of $N$ = 1. For $N$ = 3, the ionization is enhanced both during the sheath expansion at the powered electrode and during the sheath collapse at the grounded electrode. Therefore, the ion fluxes at both electrodes are increased compared to $N$ = 2. The situation is different for the neutrals, which do not react to electric fields. Their transport is dominated by diffusional motion. Since the enhanced high energy tail of the EEDF at the grounded electrode at high values of $N$ leads to higher generation rates of the neutral radicals through dissociation processes, more neutrals can be transported to both electrodes via diffusion. Therefore, the neutral radical fluxes are increased at both electrodes as a function of $N$.

As shown in table A3 in the Apendix, $CF_2$ and CF can react with the $SiO_2$ surface, which first forms a passivation layer (such as $SiO_2C_xF_{y(s)}$) and then deposits a polymer ($CF_{x(s)}$) on the passivation layer. The ion energy threshold for removing the passivation layer through ion bombardment is only 10 eV. This process works in a way that ions with energies around 10 eV first bombard the passivation layer and sequentially generate $SiOC_xF_{y(s)}$ and $X_{SiF(s)}$. Then F reacts with $X_{SiF(s)}$ and remove the passivation layer completely. The ion energy threshold for the etching of $SiO_2$ through direct ion bombardment is 70 V, which is much higher than that for the passivation layer. Thus, the generation of the passivation layer makes the etching of $SiO_2$ by low-energy ions possible. However, the formation of the $CF_{x(s)}$ polymer can block ions from interacting with the target material ($SiO_2$). The removal of the $CF_{x(s)}$ polymer consumes a high number of ions, slowing down the etching process. In addition, H can react with the $CF_{x(s)}$ polymer and form a hydrogenated polymer on top of the $CF_{x(s)}$ polymer. The hydrogenated polymer can also consume ions and cause a reduction of the etching rate. Consequently, the ratio of the ion to neutral flux, i.e. the so-called synergy value (*Syn* value), is a key parameter in plasma etching, as it controls the polymer growth, and thus, determines the etch rate.

Figure 6 shows the total ion flux as well as the sum of the $CF_2$, CF and H neutral fluxes, the *Syn* value (defined as $(\Gamma_{CF_2} + \Gamma_{CF} + \Gamma_H)/\Gamma_{i,Total}$ in this work) and the mean ion energy at the powered and grounded electrode for different $N$. For F, it generally removes hydrogenated polymers or polymers, such as the surface reaction SR10, SR11, SR44 in Appendix. table A3. Since F atoms can remove the polymers and $CF_2$, CF, H can generate polymer or hydrogenated polymer, F plays an opposite role to $CF_2$, CF, H. And therefore, the calculation of the synergy value does not include the F flux. Figure 6 (a) and (b) show that the total ion flux at the grounded electrode as well as the sum of the $CF_2$, CF, and H fluxes at both electrodes increase monotonically as a function of $N$. The total ion flux and neutral radical flux at the powered electrode are always lower than those at the grounded electrode when $N$ is 2 or 3 due to the higher ionization and dissociation reaction rates near the grounded electrode. In addition, since the neutral fluxes at the powered electrode increase fast while the ion fluxes are reduced, as shown in figure 5, the *Syn* value is higher at the powered compared to the grounded electrode for $N$ = 2 and 3. This effect is less pronounced at the grounded electrode due to the stronger increase of the ion fluxes as a function of $N$. Figure 6 (d) shows that the mean ion energy at the powered electrode is slightly increased from 129.80 eV to 132.03 eV, and 132.95 eV, while it is significantly decreased from 129.80 eV to 96.19 eV, and 87.15 eV at the grounded side. With the increasing of *N*, period-averaged sheath potential drop at powered side changes from 171.52 V to 187.73 V and 189.70 V. Meanwhile, at grounded side, it changes from 171.52 V to 127.10 V and 114.00 V. Besides, the DC self-bias voltage decreases from 0 V to -64.88 V, and -78.97 V as *N* is increased from 1 to 3. This may indicate that the negative DC self-bias voltage at the powered electrode affects the sheath potential drop on both sides, causing an increase in the potential drop on the powered side and a decrease in the potential drop on the grounded side.

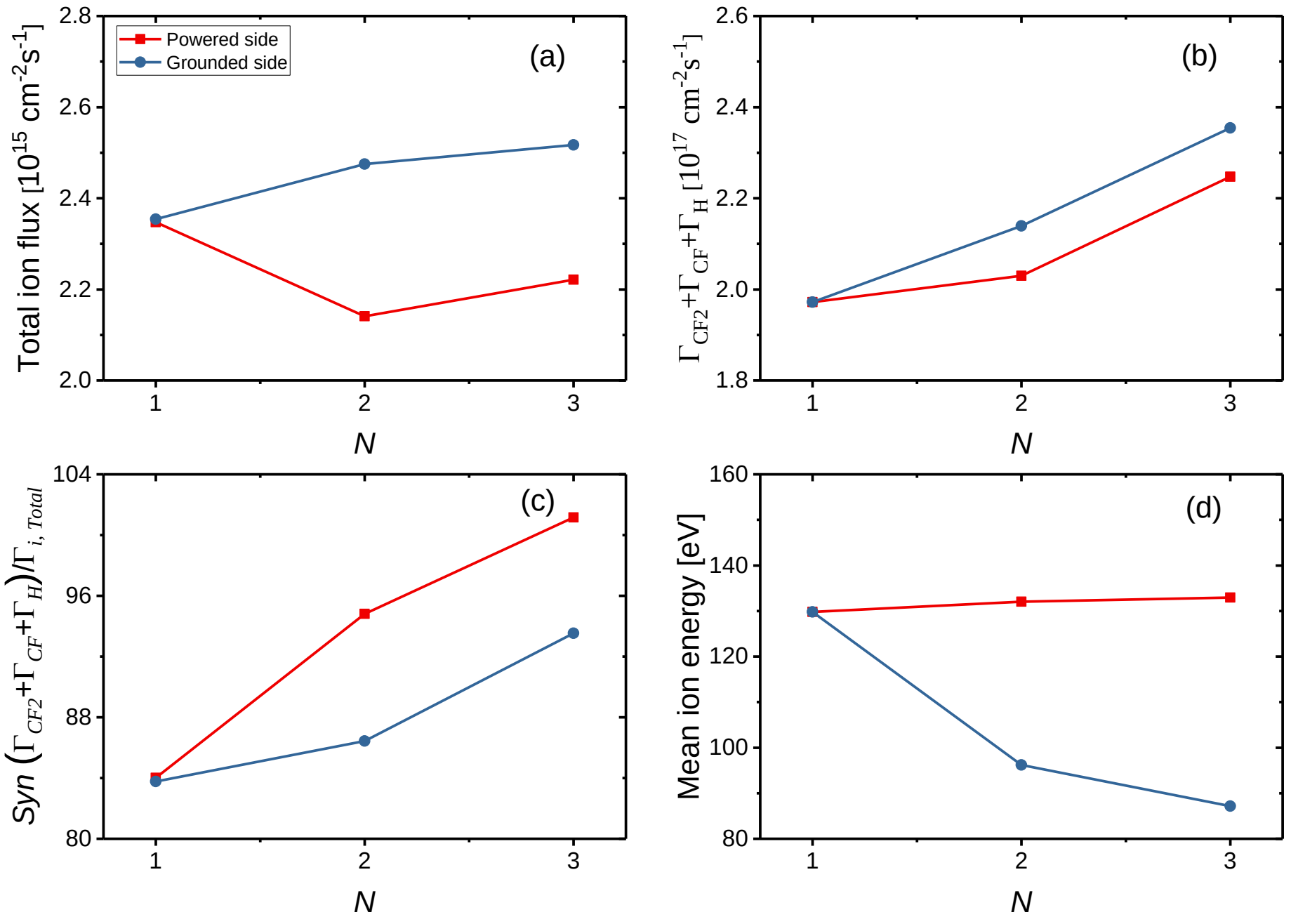

**Figure 6** Total ion flux (a), the sum of the $CF_2$, CF and H fluxes (b), the *Syn* value (defined as $(\Gamma_{CF_2} + \Gamma_{CF} + \Gamma_H)/\Gamma_{i,Total}$) (c) and the mean ion energy (d) at the powered and the grounded electrodes at *N* = 1, 2, 3. The discharge conditions are the same as those in figure 3.

Figure 7 shows the etching profiles obtained from the feature profile evolution model after a fixed etching time of 150 seconds in the cases of different *N* and at both the powered and the grounded electrodes. In the case of *N* = 2 and 3, the etch depth and, thus, the etch rate are higher at the grounded than that at the powered electrode. This is caused by the following two reasons: (i) The ion flux is higher at the grounded electrode although the mean ion energy at both electrodes is above the threshold for $SiO_2$ removal. (ii) The *Syn* value is larger at the powered electrode than that at the grounded electrode. The higher *Syn* value at the powered electrode causes more polymer deposition, which slows down the etching of the $SiO_2$. In addition, at the powered electrode, the etch depth decreases first and then increases slightly from about 74 nm to 69 nm, 71 nm as *N* is increased from 1 to 3. This is because the total ion flux is first decreased and then increased as a function of $N$, while the total neutral flux increases monotonically. At the grounded electrode, the etching depth increases slightly from about 74 nm to 75 nm, 77 nm as a function of $N$ due to the increase of the total ion flux. Furthermore, there is no significant change in the etching depth as the $N$ increases. This is related to the fact that the *Syn* value (figure 6 (c)) or neutral fluxes (figure 6 (b)) gradually increases with the increasing of the number of harmonics $N$ on both electrode sides, although the ion flux also increases in the process.

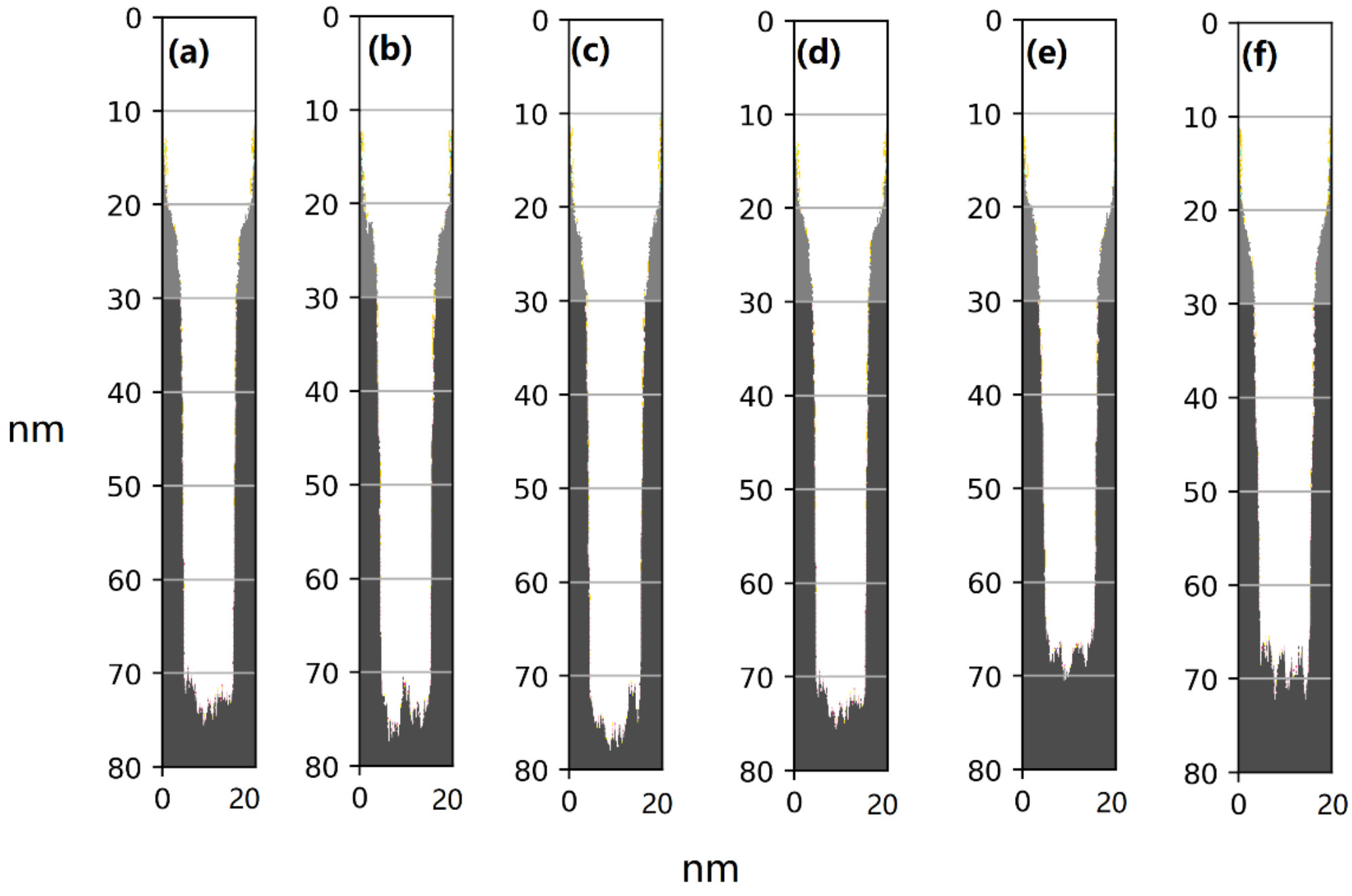

**Figure 7** Etching profiles at the grounded electrode for (a) $N = 1$, (b) $N = 2$, (c) $N = 3$ and at the powered electrode for (d) $N = 1$, (e) $N = 2$, (f) $N = 3$ after an etching time of 150 s. The discharge conditions are the same as those in figure 3. (Grey material is the photoresist, black material is $SiO_2$, other colors represent polymer or passivation layers on the material surface).

Overall, enhancing the slope asymmetry by increasing the number of harmonics of the sawtooth-up waveform can strongly change the electron power absorption dynamics, the spatio-temporal EEDF, the mean ion energy, as well as the ion and neutral fluxes at the boundary surfaces, which further affects the etching processes. Different etch profiles are formed at the powered and the grounded electrode due to the asymmetry of the discharge in the cases of $N = 2$ and $N = 3$. The neutral to ion flux ratio, the mean ion energy and thus, the etch rate can be controlled by changing $N$. Based on these simulation results, we can conclude that higher etch rates can be obtained by placing the wafer on the grounded side and set a high value of $N$, while by placing the showerhead gas inlet on the powered side can reduce the etching of the showerhead. In conclusion, the results show that TVWs provide electrical control of etch relevant plasma parameters.

### 3.2 Effects of the Ar/$CHF_3$ gas mixing ratio on the electron dynamics and etching profiles

In this section, the effects of varying the Ar/$CHF_3$ gas mixing ratio on the electron dynamics and etching results are investigated. The spatio-temporal evolutions of the electron density, $F^-$ density, electric field, electron power absorption rate, total ionization rate, as well as the temporal variations of the driving voltage waveform and the electron current density are shown in figure 8 for Ar/$CHF_3$ gas maxing ratios of 90/10, 50/50, and 10/90. A sawtooth-up waveform with $N = 2$ is used. As shown in figures 8 (a), as the $CHF_3$ gas ratio increases from 10% to 50%, the electron density in the bulk region increases, and the maximum of the electron density at the positions of the maximum sheath widths becomes more pronounced near both electrodes. The higher electron density in the bulk region in the case of 50% $CHF_3$ is mainly caused by some low-energy electron related ionization, such as $Ar^*$ ionization (table A1, R5), and collision ionization reaction between $Ar^*$ and $Ar^*$ (table A2, R31). When the $CHF_3$ gas ratio further increases to 90%, the maximum electron density near the sheath edges is almost unchanged at both electrodes, while the electron density in the bulk region is decreased compared to case of 50% $CHF_3$. The reduction of electron density in the bulk region is mainly caused by the fact that $Ar^*$ related ionization rate is attenuated as the proportion of Ar gas decreases.

The $F^-$ density in the bulk region increases as the $CHF_3$ gas admixtures increase from 10% to 50%, but decreases when the $CHF_3$ gas admixture increases to 90%. The reduction of $F^-$ density is considered to be partially induced by the reactions between $F^-$ and the $CHF_3$-associated particles, i.e. R26, R27, R28, R29, R32, R33, R34, and R35 in table A2 in the Appendix. As shown in figure 10, the $CHF_3$-associated neutral fluxes, such as $CF_2$, CF, F, all increases, which indicates that the loss rate of $F^-$ is increased, which reduces the $F^-$ density. As a result of the changes of the electron and $F^-$ density, the electronegativity increases from 8.15 to 13.57 and 13.43 as the $CHF_3$ ratio increases from 10% to 50% and 90%. As shown in figure 8, the increased electronegativity in the case of 50% $CHF_3$ and 90% $CHF_3$ leads to an enhanced drift ambipolar electric field in the bulk and near the collapsing sheath edges, which enhances the electron heating under the DA mode of the discharge. As a result, as shown in figure 8 (e1)-(e3), stronger ionization peaks are formed near the grounded electrode and the discharge asymmetry is increased. The DC self-bias increases from −51.46 V to −66.38 V and −63.87 V as the $CHF_3$ ratio increases from 10% to 50% and 90%.

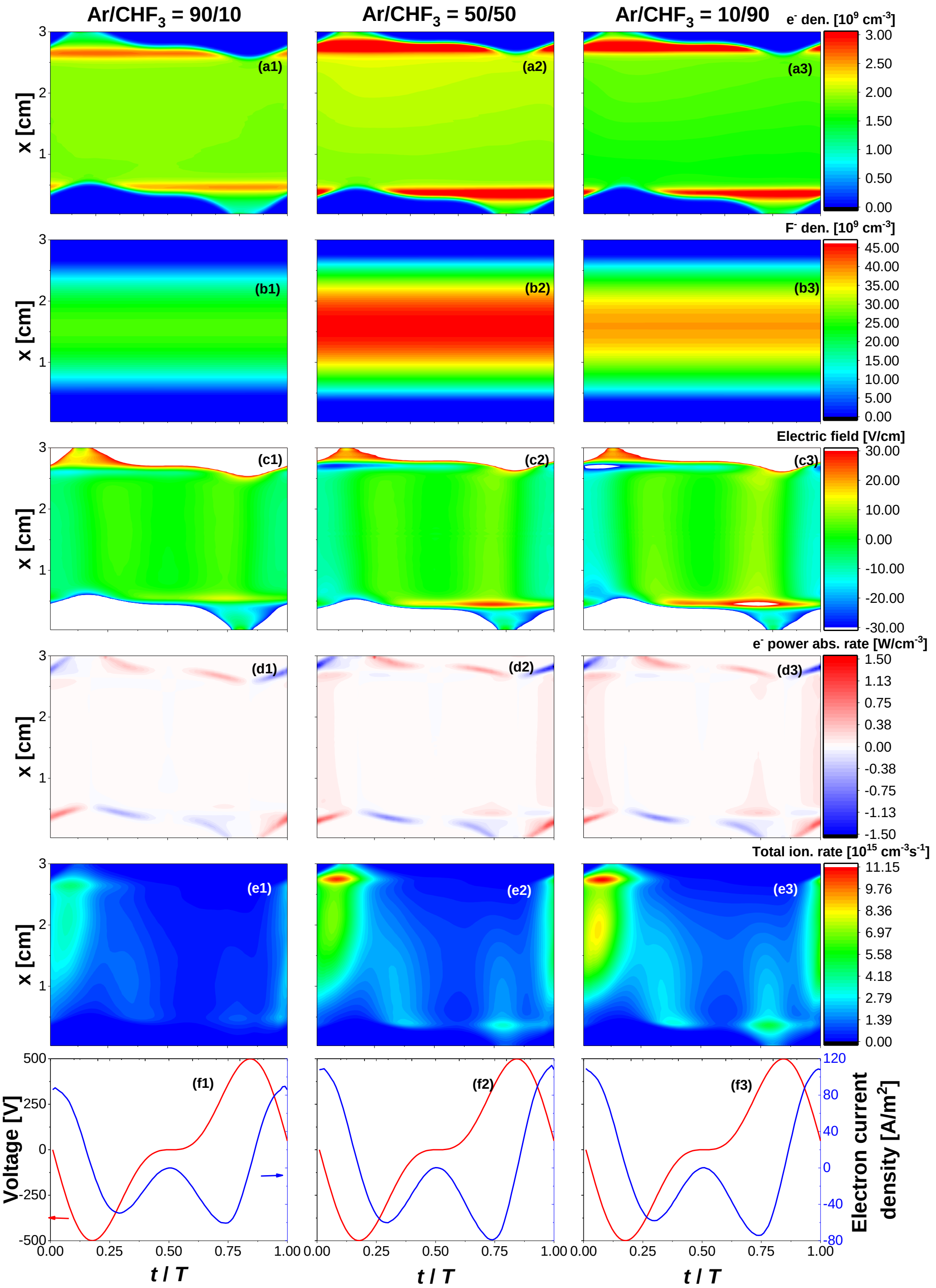

**Figure 8** Spatio-temporal evolution of the electron density (a1)-(a3), $F^-$ density (b1)-(b3), electric field (c1)-(c2), electron power absorption rate (d1)-(d3), total ionization rate (e1)-(e3), and the driving voltage waveform as well as electron current density (f1)-(f3) at 100 mTorr for different gas mixing ratios of Ar/$CHF_3$ of 90/10, 50/50, 10/90. Discharge conditions: 3 cm electrode gap, sawtooth-up voltage waveform with $N$ = 2, a fundamental frequency of 13.56 MHz and a peak-to-peak voltage of 500 V.

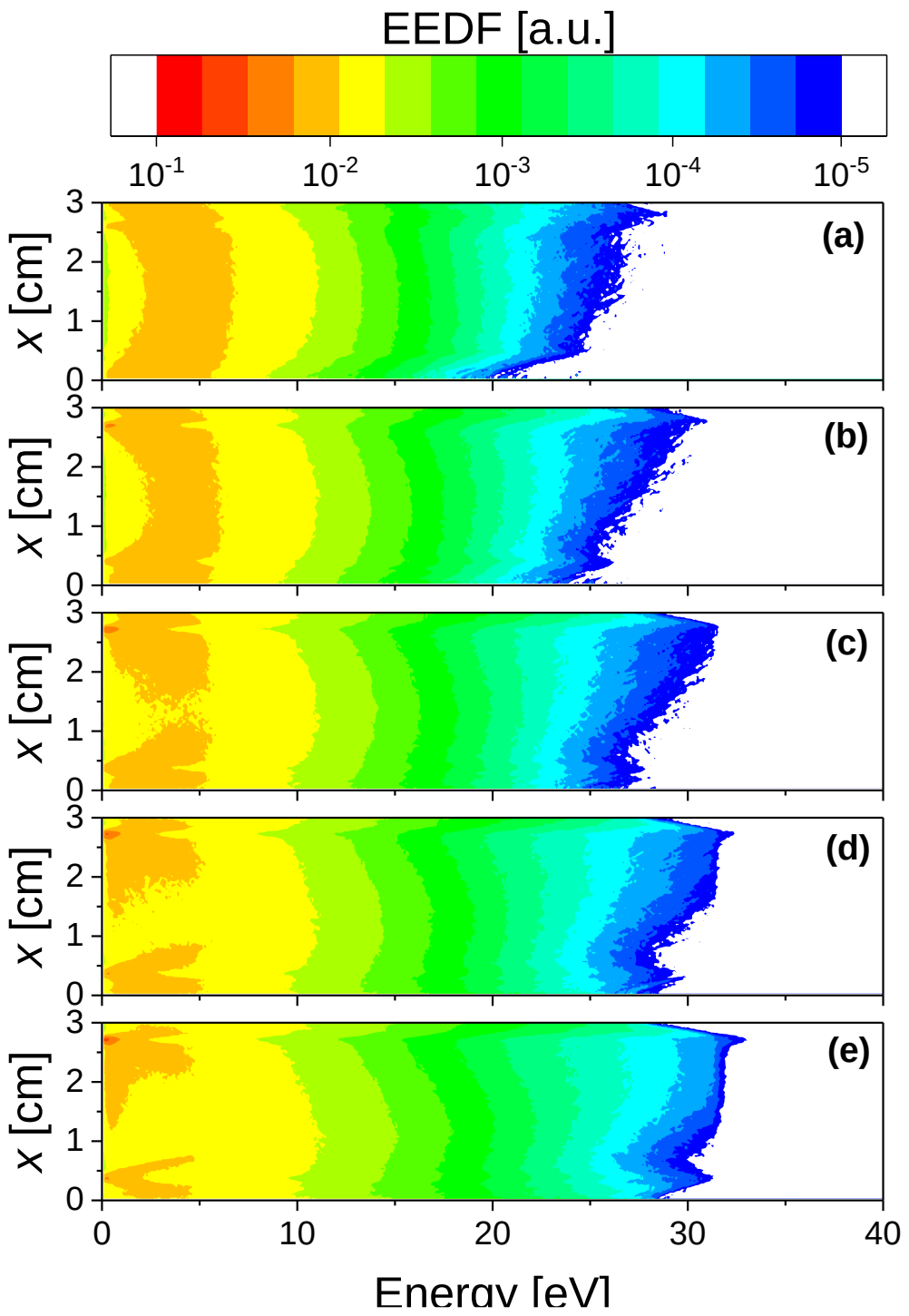

**Figure 9** Spatially resolved and time averaged EEDF for different gas mixtures: (a) Ar/$CHF_3$ = 90/10, (b) Ar/$CHF_3$ = 70/30, (c) Ar/$CHF_3$ = 50/50, (d) Ar/$CHF_3$ = 30/70, and (e) Ar/$CHF_3$ = 10/90. The other discharge conditions are the same as those in figure 8.

Figures 9 shows the spatially resolved EEDFs at different Ar/$CHF_3$ gas maxing ratios. As the $CHF_3$ admixture increases, the high energy tail of the EEDF is enhanced especially near the grounded electrode, where the maximum electron energy increases from around 28 eV to 33 eV. This is caused by the enhanced electron heating by the drift-ambipolar electric field, which significantly increases the number of high-energy electrons, and reduces the number of low-energy electrons. Due to the dependence of the corresponding cross section on the electron energy, this leads to more electron impact ionization collisions rather than attachment collisions. This is considered to be another reason for the decreased $F^-$ density in the bulk region in the case of 90% $CHF_3$.

Figures 10 shows the ion and neutral radical fluxes at both electrodes at different Ar/$CHF_3$ gas mixing ratios. As shown, except for $Ar^+$, the other ion and neutral radical fluxes increase at both electrodes as the increasing of $CHF_3$ admixture since the increased $CHF_3$ density and the enhanced high energy tail of the EEDF lead to enlarged generation rates of these particles. The $Ar^+$ flux is first increased by the enhanced Ar- and $Ar^*$-related ionization through high-energy electrons, and then decreased due to the reduced Ar density.

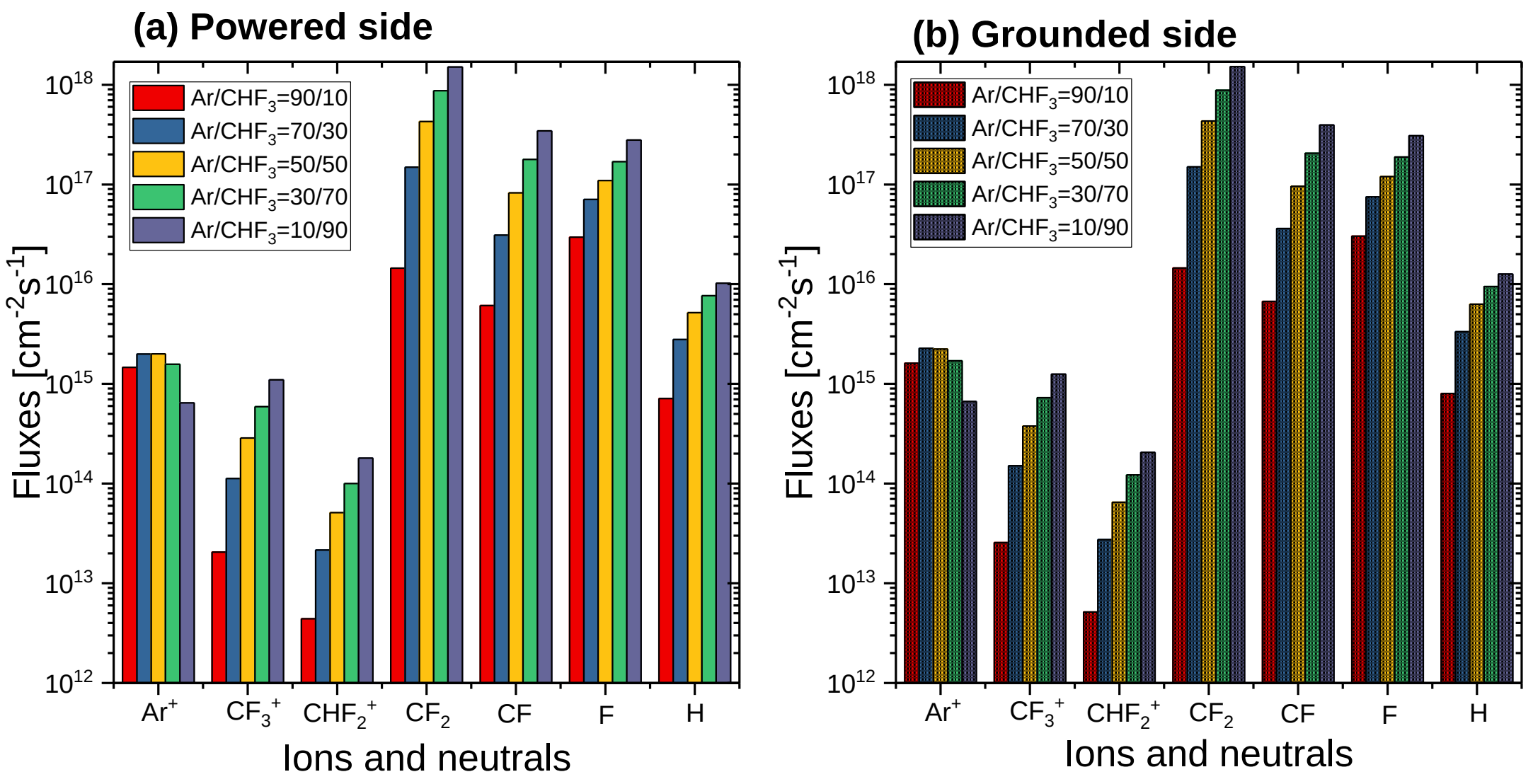

**Figure 10** $Ar^+$, $CF_3^+$, $CHF_2^+$ $CF_2$, CF, F, H fluxes at the (a) powered electrode and the (b) grounded electrode for different gas mixtures of Ar/$CHF_3$ = 90/10, 70/30, 50/50, 30/70, 10/90. The discharge conditions are the same as those in figure 8.

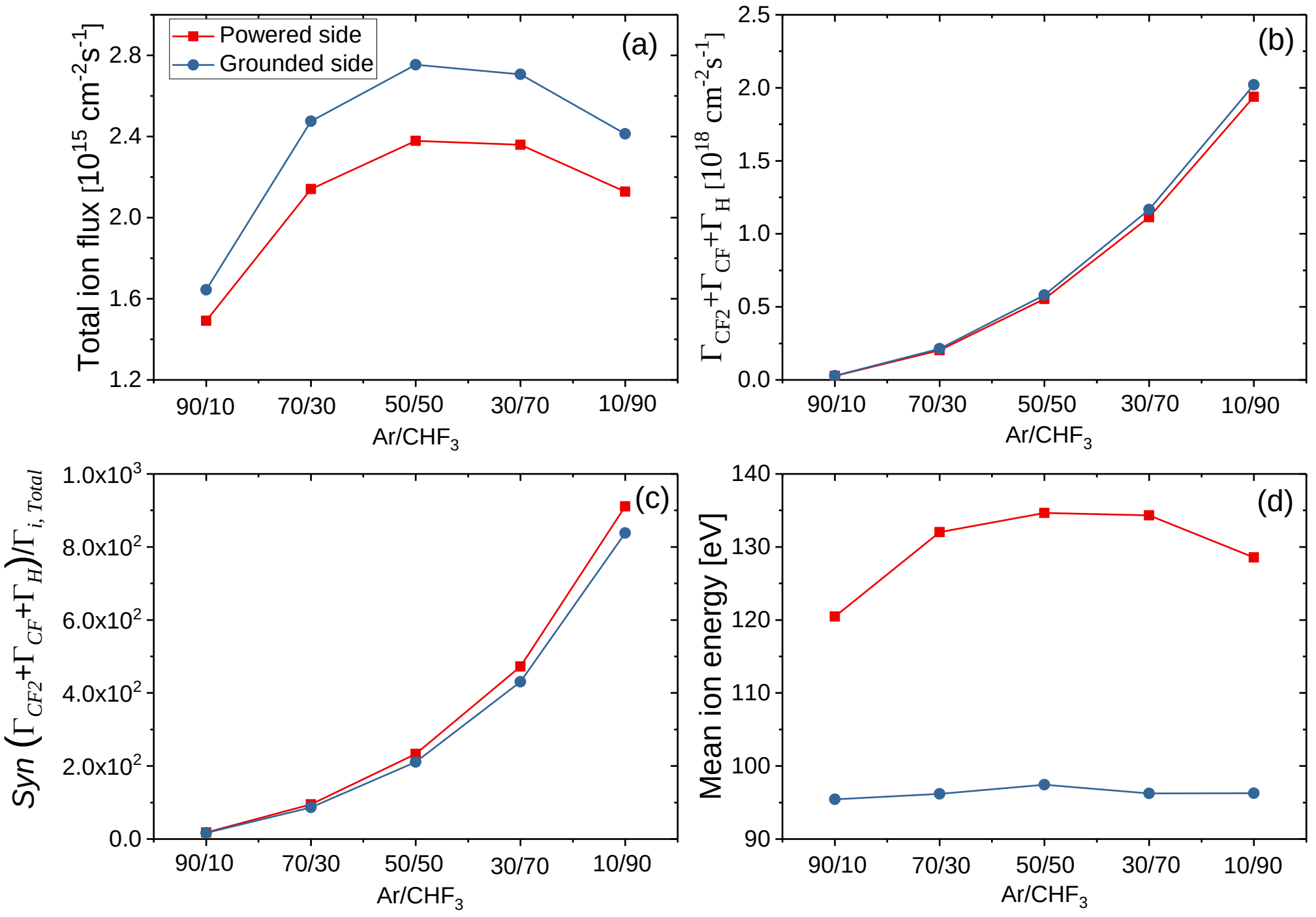

**Figure 11** Total ion flux (a), the sum of $CF_2$, CF and H flux (b), *Syn* value ($\Gamma_{i,Total}/(\Gamma_{CF_2} + \Gamma_{CF} + \Gamma_H)$) (c), and mean ion energy (d) at the powered electrode and the grounded electrode for gas mixtures of Ar/$CHF_3$ = 90/10, 70/30, 50/50, 30/70, 10/90. The discharge conditions are the same as those in figure 8.

Figure 11 presents the total ion flux, the sum of the $CF_2$, CF and H flux, the *Syn* value, and the mean ion energy as a function of the Ar/$CHF_3$ mixing ratio. At both electrodes the total ion flux first increases and then decreases as the $CHF_3$ admixture increases, while the neutral flux increases monotonically. As shown in figure 10, the total ion flux predominantly consists of $Ar^+$ ions at $CHF_3$ admixtures below 90%, while $CF_3^+$ dominate only for Ar admixture of 10% or less. Therefore, the total ion flux follows the trend of the $Ar^+$ flux for Ar admixtures above 10%. The continuous increase of the $CHF_3$ related neutral fluxes is caused by the increase of the number of $CHF_3$ molecules as potential sources of reactive radicals and the increased number of high-energy electrons. The ion and neutral fluxes are higher at the grounded electrode than that at the powered side due to the presence of more high energy electrons at the grounded electrode, as illustrated in figures 8 and 9. The *Syn* value increases significantly as the increase of the $CHF_3$ admixture at both electrodes. The mean ion energy at the powered electrode follows the trend of the DC self-bias, i.e., it first increases and then decreases as a function of the $CHF_3$ admixture. At the grounded electrode the mean ion energy remains approximately constant at 95 eV. In addition, changes in the gas ratio may also affect the ion energy as the ions collide with the background gases Ar and $CHF_3$, and there are differences in the collision cross sections for each process.

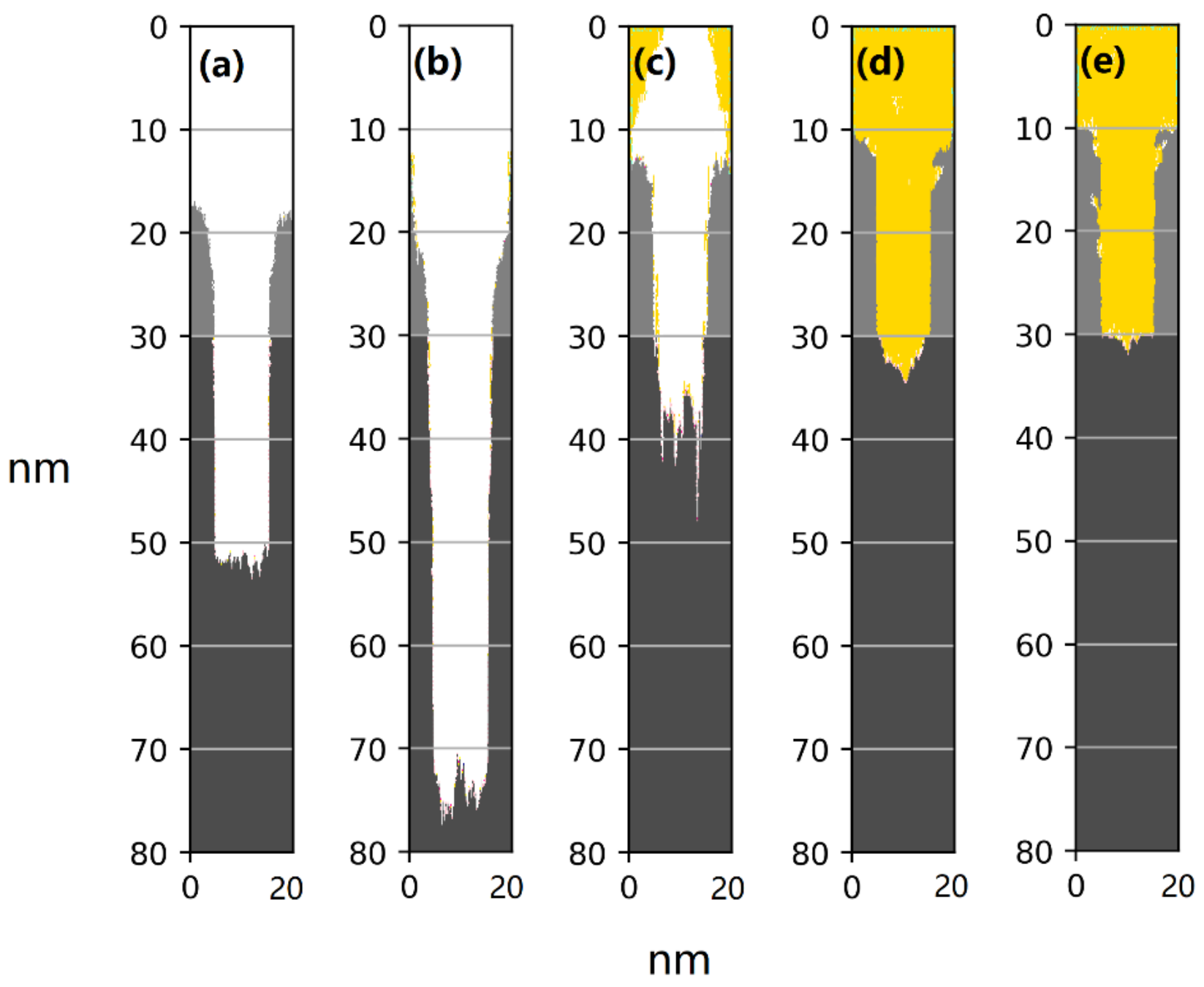

**Figure 12** Etching profiles at the grounded electrode for different gas mixtures of Ar/$CHF_3$: (a) 90/10, (b) 70/30, (c) 50/50, (d) 30/70, (e) 10/90 after an etching time of 150 s. The discharge conditions are the same as those in figure 8.

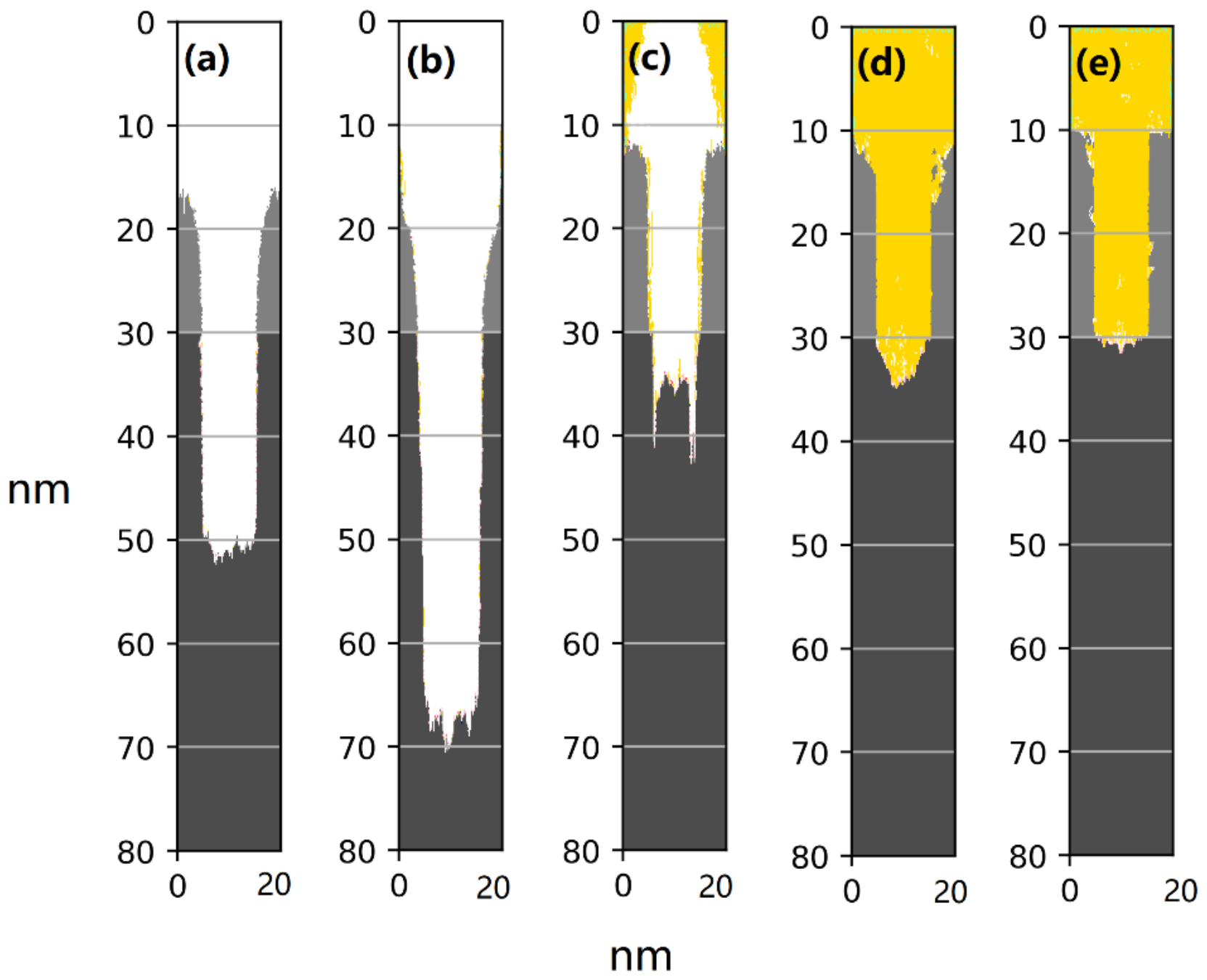

**Figure 13** Etching profiles at the powered electrode for different gas mixtures of Ar/$CHF_3$: (a) 90/10, (b) 70/30, (c) 50/50, (d) 30/70, (e) 10/90 after an etching time of 150 s. The discharge conditions are the same as those in figure 8.

Figures 12 and 13 show the etching profiles at the grounded electrode and the powered electrode at different gas mixing ratios after an etching time of 150 s. The etching depth/rate initially increases (from 10% $CHF_3$ to 30% $CHF_3$) and then decreases at both electrodes as increasing the $CHF_3$ admixture, which even becomes negative, i.e. deposition occurs and the etch feature is filled with polymer, as the $CHF_3$ ratio is above 70%. The first increased etching depth is caused by an improved surface coverage of the neutral radicals, which leads to near-complete surface passivation and a subsequent acceleration of the etching. Increasing the $CHF_3$ admixture from 30% to 50% results in a significant reduction of the etching depth due to the enhanced abundant polymer (or hydrogenated polymer) generation on the side walls and trench bottom. As the $CHF_3$ admixture further increases to 70 % and 90 %, more polymer deposition occurs, so that the trench is filled by polymer, preventing ions from reaching the surface of the target material. Since the total ion flux at the grounded electrode is higher than that at the powered electrode, etching at the grounded electrode is a bit faster, as shown in figures 12 (b), (c) and 13 (b), (c). Moreover, in the case of 50% $CHF_3$, the bottom surface of the trench is very rough at both electrodes, exhibiting a structure of “keyhole”. This is because the polymer deposited in the trench cannot be removed completely and immediately. In contrast, when the $CHF_3$ admixture is less than 30%, the bottom of the trench is relatively flat, especially at 10% $CHF_3$ admixture, although the lower neutral radical fluxes result in less passivation layer generation and a low etching rate in the case of 10% $CHF_3$. Based on these results, the optimum Ar/$CHF_3$ gas mixing ratio is 70/30 under the discharge conditions studied here, which gives a high etching rate and a relatively low surface roughness at the trench bottom. In addition, the photoresist near the notch is generally easier to remove than that near the left and right boundaries in most cases. This is due to the fact that the isotropic electrons can reach this position more easily compared to the anisotropic ions. Therefore, the potential on the photoresist surface near the notch side is typically negative, which accelerates ions and enhances the removal of the photoresist layer [70].

## 4 Conclusions

Capacitively coupled RF discharges operated in $CHF_3$/Ar and driven by slope asymmetric sawtooth tailored voltage waveforms were investigated using a one-dimensional fluid/MC model coupled with an etching profile evolution model. The effects of changing the number of consecutive harmonics, *N*, of the fundamental driving frequency of 13.56 MHz used to synthesize the tailored driving voltage waveform at constant peak-to-peak voltage and of varying the gas mixture on the electron dynamics, various plasma parameters and plama etching of patterned $SiO_2$ wafers were revealed.

As the number of harmonics *N* increases, the electron power absorption rate is enhanced at the time of maximum current in each fundamental RF period, due to the enhanced slope asymmetry of the driving voltage waveform. This leads to higher ionization rates near the grounded electrode and the generation of a negative DC self-bias. In addition, electrons are accelerated more strongly by the fast sheath expansion near the powered electrode in the multi-frequency sawtooth-up cases compared to the single frequncy case. As a result, the discharge mode is changed from the DA- to the $\alpha$-DA hybrid mode. Since the ionization and the ion fluxes are enhanced at the grounded electrode, the etching rate at the grounded electrode is higher than that at the powered electrode. The etching rate decreases at the powered electrode as the number of harmonics is increased from one to two, because the ion flux is decreased and the neutral fluxes are increased at this electrode, which leads to a large amount of polymer deposition.

By increasing the $CHF_3$ asmixture, the electronegativity is increased and the drift-ambipolar electric field is enhanced, which leads to stronger electron heating and ionization in the plasma bulk and near the grounded electrode. This contributes to a larger asymmetry of the discharge and a higher value of the DC self-bias. The enhanced electron heating also results in a larger high energy tail of the EEDF over the whole discharge gap. In addition, the neutral fluxes at the electrodes are incerased as the $CHF_3$ admixture is increased. This leads to a first increased and then decreased etch rate, since initially the increased neutral radical fluxes enhance the deposition of the passivation layer, which speeds up the etching, while the further increase of the neutral fluxes results in polymer deposition which stops the etching.

In summary, a detailed understanding of the Ar/$CHF_3$ discharge characteristics and the roles of neutral radicals and ions on the etching of $SiO_2$ was obtained under a wide range of discharge conditions in this work. TVWs were found to be able to control the neutral to ion flux ratio and therefore, control the etch rate etc. Varying the Ar/$CHF_3$ was also found to have significant effects on the discharge and the etch process. These results are expected to be highly relavent for the knowledge-based process control in etching within the semiconductor industry.

## Data availability statement

All data that support the findings of this study are included within the article (and any supplementary files).

## Acknowledgments

This work was supported by the Natural Science Foundation of China under Grant Nos. 12347131, 12405289, 12020101005, 12475202, the Liaoning Provincial Natural Science Foundation Joint Fund under Grant Nos. 2023-BSBA-089, the China Scholarship Council (No. 202306060179), and the German Research Foundation via the grant 428942393.

## Appendix. Tables

**Table A1** Electron-neutral reactions considered in this work and their corresponding energy thresholds. The rate coefficients are calculated based on cross sections taken from the references provided in the last column.

| # | Reaction | Threshold (eV) | Reference |
|---|---|---|---|
| R1 | $e^-+Ar \rightarrow Ar^++2e^-$ | 15.6 | [56] |
| R2 | $e^-+Ar \rightarrow Ar^*+e^-$ | 11.56 | [57] |
| R3 | $e^-+Ar \rightarrow Ar+e^-$ | 0 | [57] |
| R4 | $e^-+Ar^* \rightarrow Ar^++2e^-$ | 4.12 | [56] |
| R5 | $e^-+Ar^* \rightarrow Ar+e^-$ | - | [56] |
| R6 | $e^-+CHF_3 \rightarrow CHF_3+e^-$ | 0 | [58] |
| R7 | $e^-+CHF_3 \rightarrow CHF_3(v1,4)+e^-$ | 0.37 | [58] |
| R8 | $e^-+CHF_3 \rightarrow CHF_3(v2,5)+e^-$ | 0.18 | [58] |
| R9 | $e^-+CHF_3 \rightarrow CHF_3(v3,6)+e^-$ | 0.13 | [58] |
| R10 | $e^-+CHF_3 \rightarrow CHF_2+F^-$ | 0 | [58] |
| R11 | $e^-+CHF_3 \rightarrow CHF_2^++F^-+e^-$ | 11.5 | [58] |
| R12 | $e^-+CHF_3 \rightarrow CF_3+H+e^-$ | 11.0 | [58] |
| R13 | $e^-+CHF_3 \rightarrow CHF_2+F+e^-$ | 13.0 | [58] |
| R14 | $e^-+CHF_3 \rightarrow CF_2+H+F+e^-$ | 23.6 | [58] |
| R15 | $e^-+CHF_3 \rightarrow CHF+2F+e^-$ | 35.0 | [58] |
| R16 | $e^-+CHF_3 \rightarrow CF+H+2F+e^-$ | 19.5 | [58] |
| R17 | $e^-+CHF_3 \rightarrow CF+H+F_2+e^-$ | 19.5 | [58] |
| R18 | $e^-+CHF_3 \rightarrow CF_3+H+e^-$ | 11.0 | [58] |
| R19 | $e^-+CHF_3 \rightarrow CF_3^++H+2e^-$ | 15.2 | [58] |
| R20 | $e^-+CHF_3 \rightarrow CHF_2^++F+2e^-$ | 16.8 | [58] |
| R21 | $e^-$+$CHF_3$ → $CF_2^+$+HF+2$e^-$ | 17.6 | [58] |
| R22 | $e^-$+$CHF_3$ → $CHF^+$+2F+2$e^-$ | 19.8 | [58] |
| R23 | $e^-$+$CHF_3$ → $CF^+$+HF+F+2$e^-$ | 20.9 | [58] |
| R24 | $e^-$+$CHF_3$ → $CH^+$+$F_2$+F+2$e^-$ | 33.5 | [58] |
| R25 | $e^-$+$CHF_3$ → $F^+$+$CHF_2$+2$e^-$ | 37.0 | [58] |

**Table A2** Ion-ion, neutral-neutral, ion-neutral, electron-ion reactions and their corresponding rate coefficients. The last column shows the references from which the rate coefficients are taken from.

| # | Reaction | Rate coefficient ($cm^3/s$) | Reference |
|---|---|---|---|
| R26 | $F^-$+$Ar^+$ → Ar+F | $2.0\times10^{-7}$ | [59] |
| R27 | $F^-$+$CF_2^+$ → $CF_2$+F | $9.1\times10^{-8}$ | [59] |
| R28 | $F^-$+$CF_3^+$ → $CF_3$+F | $1.0\times10^{-7}$ | [59] |
| R29 | $F^-$+$CF^+$ → CF+F | $9.8\times10^{-8}$ | [59] |
| R30 | $F^-$+$CHF_2^+$ → $CHF_2$+F | $1.0\times10^{-7}$ | [59] |
| R31 | $Ar^*$+$Ar^*$ → Ar+$Ar^+$+$e^-$ | $5.0\times10^{-10}$ | [60] |
| R32 | $F^-$+$CF_3$→$CF_4$+$e^-$ | $5.0\times10^{-10}$ | [55] |
| R33 | $F^-$+$CF_2$→$CF_3$+$e^-$ | $5.0\times10^{-10}$ | [55] |
| R34 | $F^-$+CF→$CF_2$+$e^-$ | $5.0\times10^{-10}$ | [55] |
| R35 | $F^-$+F→$F_2$+$e^-$ | $1.0\times10^{-10}$ | [55] |
| R36 | e+$CF_3^+$→$CF_2$+F | $7.185\times10^{-9}\times Te^{-0.6311}\exp(-0.056/Te)$ | [61] |
| R37 | e+$CHF_2^+$→$CF_2$+H | $7.185\times10^{-9}\times Te^{-0.6311}\exp(-0.056/Te)$ | [61] |
| R38 | e+$CHF^+$→CF+H | $7.185\times10^{-9}\times Te^{-0.6311}\exp(-0.056/Te)$ | [61] |
| R39 | e+$CF_2^+$→CF+F | $7.185\times10^{-9}\times Te^{-0.6311}\exp(-0.056/Te)$ | [61] |
| R40 | e+$CF^+$→C+F | $7.185\times10^{-9}\times Te^{-0.6311}\exp(-0.056/Te)$ | [61] |
| R41 | e+$Ar^+$→Ar | $8.15\times10^{-13}$ $Te^{-0.5}$ | [59] |
| R42 | $CF^+$+$CF_2$→$CF_2^+$+CF | $1.0\times10^{-9}$ | [55] |
| R43 | $CF_3^+$+$CF_3$→$CF_3$+ $CF_3^+$ | $1.0\times10^{-9}$ | [55] |
| R44 | $CF^+$+$CF_3$→CF+$CF_3^+$ | $1.71\times10^{-9}$ | [55] |
| R45 | $F^+$+$CF_3$→$F_2$+$CF_2^+$ | $2.9\times10^{-9}$ | [55] |
| R46 | $CF^+ + CF \rightarrow CF + CF^+$ | $1.0\times10^{-9}$ | [55] |
| R47 | $Ar^* + M \rightarrow Ar + M$ | $3.0\times10^{-15}$ | [62] |
| R48 | $Ar^* + M + M \rightarrow Ar + M + M$ | $1.1\times10^{-31}$ | [62] |
| R49 | $CF_3 + CHF_2 \rightarrow CHF_3 + CF_2$ | $1.7\times10^{-12}$ | [59] |
| R50 | $CF + F \rightarrow CF_2$ | $1.4\times10^{-15}$ | [55] |
| R51 | $CF_2 + F \rightarrow CF_3$ | $1.25\times10^{-13}$ | [55] |
| R52 | $CHF + F \rightarrow CF + HF$ | $5.0\times10^{-11}$ | [59] |
| R53 | $CHF_2 + F \rightarrow CF_2 + HF$ | $5.0\times10^{-11}$ | [59] |
| R54 | $CHF_3 + F \rightarrow CF_3 + HF$ | $1.5\times10^{-13}$ | [59] |
| R55 | $F + F \rightarrow F_2$ | $1.2\times10^{-17}$ | [59] |
| R56 | $CF_2 + F_2 \rightarrow CF_3 + F$ | $2.0\times10^{-15}$ | [59] |
| R57 | $CF_2 + H \rightarrow CF + HF$ | $4.1\times10^{-11}$ | [59] |
| R58 | $CF_3 + H \rightarrow CF_2 + HF$ | $8.9\times10^{-11}$ | [55] |
| R59 | $CHF_2 + H \rightarrow CHF + HF$ | $1.1\times10^{-10}$ | [59] |
| R60 | $F_2 + H \rightarrow F + HF$ | $1.8\times10^{-12}$ | [59] |
| R61 | $CF_3 + CF \rightarrow 2CF_2$ | $5.0\times10^{-11}$ | [47] |
| R62 | $F_2 + CF \rightarrow CF_2 + F$ | $3.98\times10^{-12}$ | [63] |
| R63 | $CF + CF_2 \rightarrow C_2F_3$ | $1\times10^{-12}$ | [55] |
| R64 | $CF_2 + CF_2 \rightarrow C_2F_4$ | $4\times10^{-14}$ | [55] |
| R65 | $CF_2 + CF_3 \rightarrow C_2F_5$ | $8\times10^{-13}$ | [55] |
| R66 | $CF_3 + CF_3 \rightarrow C_2F_6$ | $7.2\times10^{-12}$ | [55] |
| R67 | $CF_3 + F \rightarrow CF_4$ | $8.5\times10^{-12}$ | [55] |
| R68 | $CF + H \rightarrow CH + F$ | $1.9\times10^{-11}$ | [55] |
| R69 | $CHF_2 + CF_2 \rightarrow C_2F_4 + H$ | $3.3\times10^{-12}$ | [55] |
| R70 | $CHF_2 + CF_2 \rightarrow CHFCF_2 + F$ | $6.6\times10^{-12}$ | [55] |
| R71 | $CHF_2 + CF_3 \rightarrow C_2HF_5$ | $2.2\times10^{-11}$ | [55] |
| R72 | $CHF_2 + CHF_2 \rightarrow CHF_2CHF_2$ | $2.9\times10^{-11}$ | [55] |
| R73 | $CHF_2 + CHF \rightarrow CHFCHF + F$ | $6.6\times10^{-12}$ | [55] |
| R74 | $CHF_2 + CHF \rightarrow CHFCF_2 + H$ | $3.3\times10^{-12}$ | [55] |
| R75 | $CHF + CF_2 \rightarrow C_2F_2 + HF$ | $1.5\times10^{-11}$ | [55] |
| R76 | $CHF + CF_2 \rightarrow CHFCF_2$ | $1.7\times10^{-11}$ | [55] |
| R77 | $CHF + CF_3 \rightarrow CHFCF_2 + F$ | $1.0\times10^{-11}$ | [55] |
| R78 | $CHF + CHF \rightarrow C_2HF + HF$ | $2.9\times10^{-11}$ | [55] |
| R79 | $CHF + CHF \rightarrow CHFCHF$ | $8.6\times10^{-12}$ | [55] |
| R80 | $CHF_2 + H \rightarrow CHF + HF$ | $1.1\times10^{-10}$ | [55] |
| R81 | $CHF+H \rightarrow CH+HF$ | $4.9 \times 10^{-10}$ | [55] |
| R82 | $CHF+HF \rightarrow CH_2F_2$ | $1.9 \times 10^{-13}$ | [55] |
| R83 | $CF_3+F_2 \rightarrow F+CF_4$ | $7 \times 10^{-14}$ | [55] |

Note: M indicates arbitrary neutrals.

**Table A3** Surface reactions considered in the model.

| # | Deposition Reactions | Reference |
|---|---|---|
| SR1 | $Si_{(s)}+CF_{2(g)} \rightarrow Si_{(s)}+CF_{x(s)}$ | [67] |
| SR2 | $SiO_{2(s)}+CF_{2(g)} \rightarrow SiO_2C_xF_{y(s)}$ | [67] |
| SR3 | $Si_3N_{4(s)}+CF_{2(g)} \rightarrow SiNC_xF_{y(s)}$ | [67] |
| SR4 | $SiO_2C_xF_{y(s)}+CF_{2(g)} \rightarrow SiO_2C_xF_{y(s)}+CF_{x(s)}$ | [67] |
| SR5 | $SiOC_xF_{y(s)}+CF_{2(g)} \rightarrow SiOC_xF_{y(s)}+CF_{x(s)}$ | [67] |
| SR6 | $SiNC_xF_{y(s)}+CF_{2(g)} \rightarrow SiNC_xF_{y(s)}+CF_{x(s)}$ | [67] |
| SR7 | $CF_{x(s)}+CF_{2(g)} \rightarrow CF_{x(s)}+CF_{x(s)}$ | [67] |
| SR8 | $CF_x{}^*{}_{(s)}+CF_{2(g)} \rightarrow CF_{x(s)}+CF_{x(s)}$ | [67] |
| SR9 | $CF_{x(s)}+CF^*{}_{(implant)} \rightarrow CF_{x(s)}+CF_{x(s)}$ | [67] |
| | Polymer Loss Reactions | |
| SR10 | $CF_{x(s)}+F_{(g)} \rightarrow CF_{4(g)}$ | [67] |
| SR11 | $CF_x{}^*{}_{(s)}+F_{(g)} \rightarrow CF_{4(g)}$ | [67] |
| SR12 | $CF_{x(s)}+Ar^+{}_{(g)} \rightarrow CF_{x(g)}+Ar$ | [67] |
| SR12 | $CF_{x(s)}+Ar^+{}_{(g)} \rightarrow CF_{x(g)}+Ar$ | [67] |
| SR13 | $CF_{x(s)}+C_xF_y{}^+{}_{(g)} \rightarrow CF_{x(g)}+CF_{2(g)}$ | [67] |
| SR14 | $SiO_{2(s)}+CF_{x(s)}+M^+ \rightarrow SiO_2C_xF_{y(s)}$ | [67] |
| | Polymer Activation | |
| SR15 | $CF_{x(s)}+Ar^+{}_{(g)} \rightarrow CF_x{}^*{}_{(s)}+Ar_{(g)}$ | [67] |
| SR16 | $CF_{x(s)}+C_xF_y{}^+{}_{(g)} \rightarrow CF_x{}^*{}_{(s)}+F^+{}_{(implant)}+C_xF_{y(g)}$ | [67] |
| SR17 | $CF_{x(s)}+C_xF_y{}^+{}_{(g)} \rightarrow CF_x{}^*{}_{(s)}+CF^*{}_{(implant)}+F^+{}_{(implant)}$ | [67] |
| SR18 | $CF_{x(s)}+C_xF_y{}^+{}_{(g)} \rightarrow CF_{x(s)}+F^+{}_{(implant)}+C_xF_{y(g)}$ | [67] |
| SR19 | $CF_{x(s)}+C_xF_y{}^+{}_{(g)} \rightarrow CF_{x(s)}+F^+{}_{(implant)}+CF_{(2)(g)}$ | [67] |
| | Silicon Reactions | |
| SR20 | $Si_{(s)}+F_{(g)}\rightarrow SiF_{(s)}$ | [67] |
| SR21 | $SiF_{(s)}+F_{(g)}\rightarrow SiF_{2(s)}$ | [67] |
| SR22 | $SiF_{2(s)}+F_{(g)}\rightarrow SiF_{3(s)}$ | [67] |
| SR23 | $SiF_{3(s)}+F_{(g)}\rightarrow SiF_{2(s)}+F_{2(g)}$ | [67] |
| SR24 | $SiF_{3(s)}+F_{(g)}\rightarrow SiF_{4(g)}$ | [67] |
| SR25 | $X_{SiF(s)}+F_{(g)}\rightarrow SiF_{4(g)}$ | [67] |
| SR26 | $Si_{(s)}+M^{+}_{(g)}\rightarrow Si_{(g)}+M_{(g)}$ | [67] |
| SR27 | $SiF_{x(s)}+M^{+}_{(g)}\rightarrow SiF_{x(g)}+M_{(g)}$ | [67] |
| SR28 | $SiF_{x(s)}+M^{+}_{(g)}\rightarrow SiF_{x(g)}+M_{(g)}$ | [67] |
| | Silicon Dioxide Reactions | |
| SR29 | $SiO_{2(s)}+M^{+}_{(g)}\rightarrow SiO_{2(g)}+M_{(g)}$ | [67] |
| SR30 | $SiO_{2(s)}+C_xF_y{}^{+}_{(g)}\rightarrow SiO_{2(s)}+CF_{(2)(g)}$ | [67] |
| SR31 | $SiO_{2(s)}+C_xF_y{}^{+}_{(g)}\rightarrow SiO_2C_xF_{y(s)}$ | [67] |
| SR32 | $SiO_2C_xF_{y(s)}+F_{(g)}\rightarrow SiO_2C_xF_{y(s)}+F_{2(g)}$ | [67] |
| SR33 | $SiO_2C_xF_{y(s)}+C_xF_y{}^{+}_{(g)}\rightarrow SiO_2C_xF_{y(s)}+CF_{(2)(g)}$ | [67] |
| SR34 | $SiO_2C_xF_{y(s)}+C_xF_y{}^{+}_{(g)}\rightarrow SiO_2C_xF_{y(s)}+CF_{x(s)}$ | [67] |
| SR35 | $SiO_2C_xF_{y(s)}+M^{+}_{(g)}\rightarrow SiOC_xF_{y(s)}+CO(F_2)_{(g)}+M_{(g)}$ | [67] |
| SR36 | $SiO_2C_xF_{y(s)}+M^{+}_{(g)}\rightarrow SiF_{4(g)}+CO(F_2)_{(g)}+M_{(g)}$ | [67] |
| SR37 | $SiO_2C_xF_{y(s)}+M^{+}_{(g)}\rightarrow SiF_{4(g)}+CO(F_2)_{(g)}+M_{(g)}$ | [67] |
| SR38 | $SiOC_xF_{y(s)}+C_xF_y{}^{+}_{(g)}\rightarrow SiOC_xF_{y(s)}+CF_{(2)(g)}$ | [67] |
| SR39 | $SiOC_xF_{y(s)}+C_xF_y{}^{+}_{(g)}\rightarrow SiOC_xF_{y(s)}+CF_{x(s)}$ | [67] |
| SR40 | $SiOC_xF_{y(s)}+M^{+}_{(g)}\rightarrow X_{SiF(s)}+CO(F_2)_{(g)}+M_{(g)}$ | [67] |
| SR41 | $SiOC_xF_{y(s)}+M^{+}_{(g)}\rightarrow SiF_{4(g)}+CO(F_2)_{(g)}+M_{(g)}$ | [67] |
| SR42 | $SiOC_xF_{y(s)}+M^{+}_{(g)}\rightarrow SiF_{4(g)}+CO(F_2)_{(g)}+M_{(g)}$ | [67] |
| | Reactions of hydrogenated polymer | |
| SR43 | $CF_{x(s)}+H_{(g)}\rightarrow CF_{x(s)}+HP_{(s)}$ | [54] |
| SR44 | $HP_{(s)}+F_{(g)}\rightarrow CF_{2(g)}$ | [54] |
| SR45 | $HP_{(s)}+H_{(g)}\rightarrow HP_{(s)}+HP_{(s)}$ | [54] |
| SR46 | $HP_{(s)}+CF_{2(g)}\rightarrow HP_{(s)}+HP_{(s)}$ | [54] |
| SR47 | $HP_{(s)}+ M^{+}_{(g)} \rightarrow CF_{2(g)}+CF_{3(h)}$ | [54] |
| | Reactions on photoresist surface | |
| S48 | $R_{(s)} + M^{+}_{(g)} \rightarrow R_{(g)} + I(h)$ | [68] |
| S49 | $R_{(s)} +CF_{2(g)} \rightarrow R_{(s)}+CF_{x(s)}$ | [68] |

Note:

1.$M^+$ represents all positive ions considered in the model, $C_xF_y{}^+_{(g)}$ are ions related to $CHF_3$, and $CF_{2(g)}$ represents the neutral particles, $CF_2$, CF generated during the Ar/$CHF_3$ discharges.

2.In addition to Si and $SiO_2$, $CF_{x(s)}$, $SiO_2C_xF_{y(s)}$ (and $SiOC_xF_{y(s)}$, $X_{SiF(s)}$), $CF_x{}^*{}_{(s)}$, $SiF_{(s)}$ (and $SiF_{2(s)}$, $SiF_{3(s)}$), $HP_{(s)}$, $R_{(s)}$ are also considered in this model, which represent fluoropolymer, passivation layer of $SiO_2$, ion activated polymer, passivation layer of Si, hydrogenated polymer, and photoresist, respectively.

## Appendix. Benchmark

Based on the results measured by experiment [55] or calculated by our simulation, figure A1 presents electron density at voltages of 90 V and 110 V, as well as radical densities at 90 V. In this part, the discharge conditions are obtain form literature [55], eg. $CHF_3$/Ar gas mixture (50/50) discharges is operated at 150 mTorr, an electrode gap of 4 cm, and a single frequency of 40 MHz, at voltages of 90 V and 110 V. Additionally, all the coefficients, such as the sticking coefficients of neutrals, the secondary electron emission coefficient, etc., are consistent with the settings of the literature [55]. As shown in figure A1, the electron and radical densities are of similar magnitude, with the neutral radical densities exhibiting consistent variation, compared with results of literature [55] and simulation results. Additionally, the measured electronegativity in the experiment ranges from approximately 3 to 6 at 90 V and from 2.5 to 5.4 at 110 V, while the calculated electronegativities are 3.76 and 3.20 for 90 V and 110 V, respectively. This indicates a consistent trend in electronegativity, decreasing with an increase in voltage. These results confirm the reliability of the model used in this study.

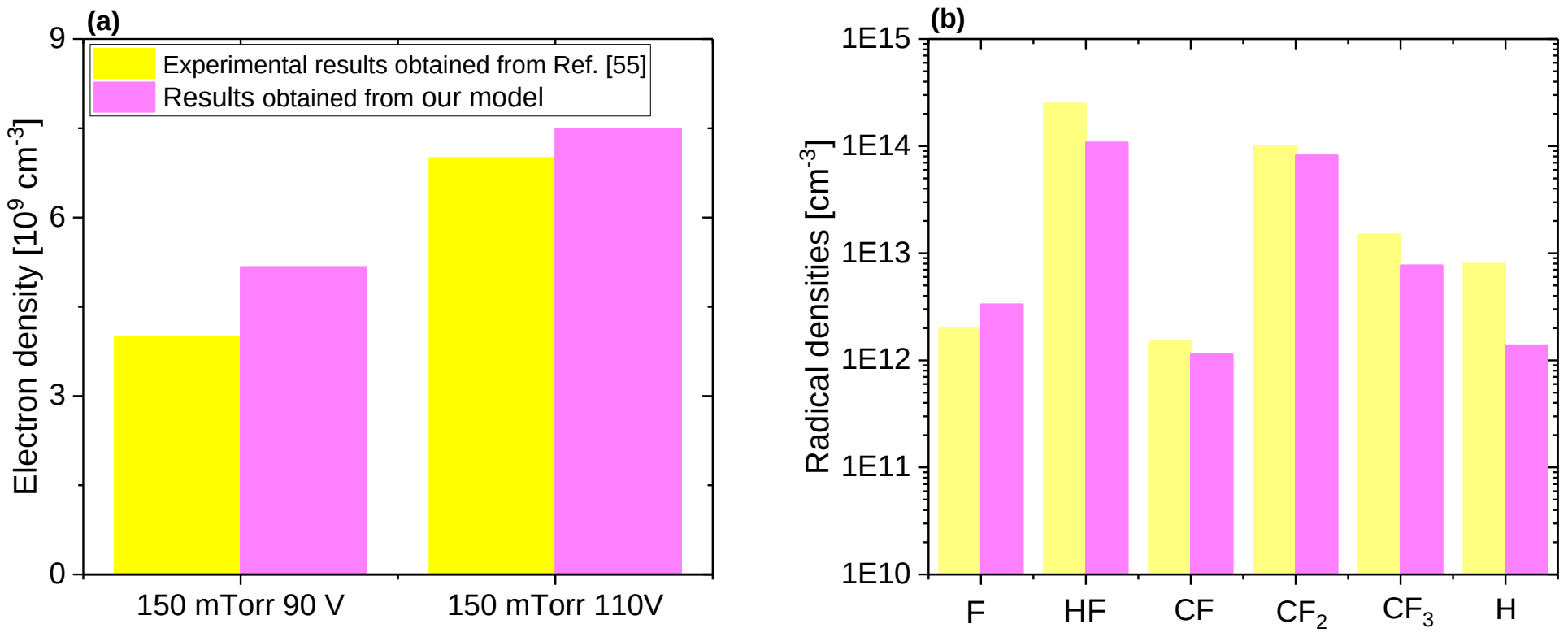

Figure A Electron density (a) and radical densities obtained from literature [55] and calculated by the model used in this work. Discharge conditions: $CHF_3$/Ar gas mixture (50/50) discharges is operated at 150 mTorr, an electrode gap of 4 cm, and a single frequency of 40 MHz, at voltages of 90 V and 110 V.

## ORCID iDs

Wan Dong https://orcid.org/0000-0002-0724-1697
Liu-Qin Song https://orcid.org/0009-0007-2327-4484
Yi-Fan Zhang https://orcid.org/0000-0001-7540-0891
Li Wang https://orcid.org/0000-0002-3106-2779
Yuan-Hong Song https://orcid.org/0000-0001-5712-9241
Julian Schulze https://orcid.org/0000-0001-7929-5734